\documentclass[12pt]{article}
\usepackage{ytableau}
\usepackage{amssymb, amsmath,amsfonts,amsthm,euscript,mathrsfs,  MnSymbol,verbatim, enumerate,multirow,bbding, slashed, multicol,
color,array}
\usepackage{tikz, tikz-cd}
\usepackage{mathdots}
\usepackage{savesym}
\usepackage{cite}
\usepackage{wasysym}
\usepackage{amscd}
\usepackage{graphicx}
\usepackage{pifont}
\usepackage{float}
\usepackage{subfig}
\usepackage[all]{xy}
\usepackage{picture}
\savesymbol{iint}
\savesymbol{iiint}
\usepackage{amsmath}
\restoresymbol{TXF}{iint}
\restoresymbol{TXF}{iiint}
\usepackage{color}
\usepackage{clay}
\usepackage{hyperref}
\usepackage{slashed}
\hypersetup{colorlinks=true}
\hypersetup{linkcolor=black}
\hypersetup{citecolor=black}
\hypersetup{urlcolor=black}
\usepackage{setspace}
\usepackage{multirow}
\usepackage{wrapfig}
\numberwithin{equation}{section}
\newcommand{\Tr}{{\rm Tr\,}}

\begin{document}

\date{}

\institution{math}{\centerline{${}^{1}$Department of Mathematics, Harvard University, Cambridge, MA, USA}}
\institution{HarvardU}{\centerline{${}^{2}$Jefferson Physical Laboratory, Harvard University, Cambridge, MA, USA}}

\title{M-theory on Elliptic Calabi-Yau Threefolds and 6$d$ Anomalies}

\authors{Mboyo Esole\worksat{\math,\HarvardU}\footnote{e-mail: {\tt esole@math.harvard.edu}}  and Shu-Heng Shao\worksat{\HarvardU}\footnote{e-mail: {\tt shshao@physics.harvard.edu}} }

\abstract{We consider the 8-supercharge 5$d$  $su(N)$ gauge theories  from M-theory compactified on  elliptic Calabi-Yau threefolds. By matching the triple intersection numbers in the elliptic Calabi-Yau with the 5$d$ Chern-Simons levels, we determine the charged matter contents for these theories. We show that all these 5$d$ theories can be lifted to 6$d$ $\mathcal{N}=(1,0)$ theories  while satisfying the anomaly cancellation equations. This suggests that the 5$d$ theories obtained from M-theory compactified on elliptic Calabi-Yau threefolds have a natural 12$d$  description, which as we know is F-theory. Furthermore, we  compute the Euler characteristics of the I$_N^s$ elliptic Calabi-Yau threefolds. }

\maketitle

\begin{spacing}{.999}
\tableofcontents
\end{spacing}
\clearpage

\section{Introduction}

 In the past two decades, a vast landscape of string vacua were explored, and various dualities between geometry and field theory were established. F-theory \cite{Vafa:1996xn,Morrison:1996na,Morrison:1996pp} and M-theory \cite{Witten:1995ex} are two parent theories that elegantly unify  all known string theories. M-theory  on $\mathscr{T} \times \mathbb{R}^{1,4}$ is expected to be dual to F-theory on $\mathscr{T}  \times S^1\times \mathbb{R}^{1,4}$, where $\mathscr{T}$ is an elliptically fibered Calabi-Yau threefold \cite{Denef:2008wq}. In this paper, we approach the F/M-theory duality from the following point of view: we start by only assuming the existence of M-theory, and study in details the low energy 5$d$ theories from M-theory compactified on elliptic Calabi-Yau threefolds.  Their F-theory origins will only emerge when the dust  settles.

Specifically, we consider  Calabi-Yau threefolds that are elliptically fibered over some algebraic surfaces $B$. The base $B$ can either be compact or non-compact. Over a generic point on the base $B$, the fiber is a smooth elliptic curve but  becomes singular over   codimension one loci in $B$. As the simplest example, we focus on the case where the singular fiber is of the type I$_N^s$, supported on a single irreducible nonsingular curve $E_0$ inside $B$. The I$_N^s$ singular fiber  consists of $N$ copies of $\mathbb{P}^1$ intersecting as an affine $su(N)$ Dynkin diagram. We will refer to these elliptic Calabi-Yau threefolds as the I$_N^s$ models.

\begin{figure}[h!]
\begin{center}
\includegraphics[width=.45\textwidth]{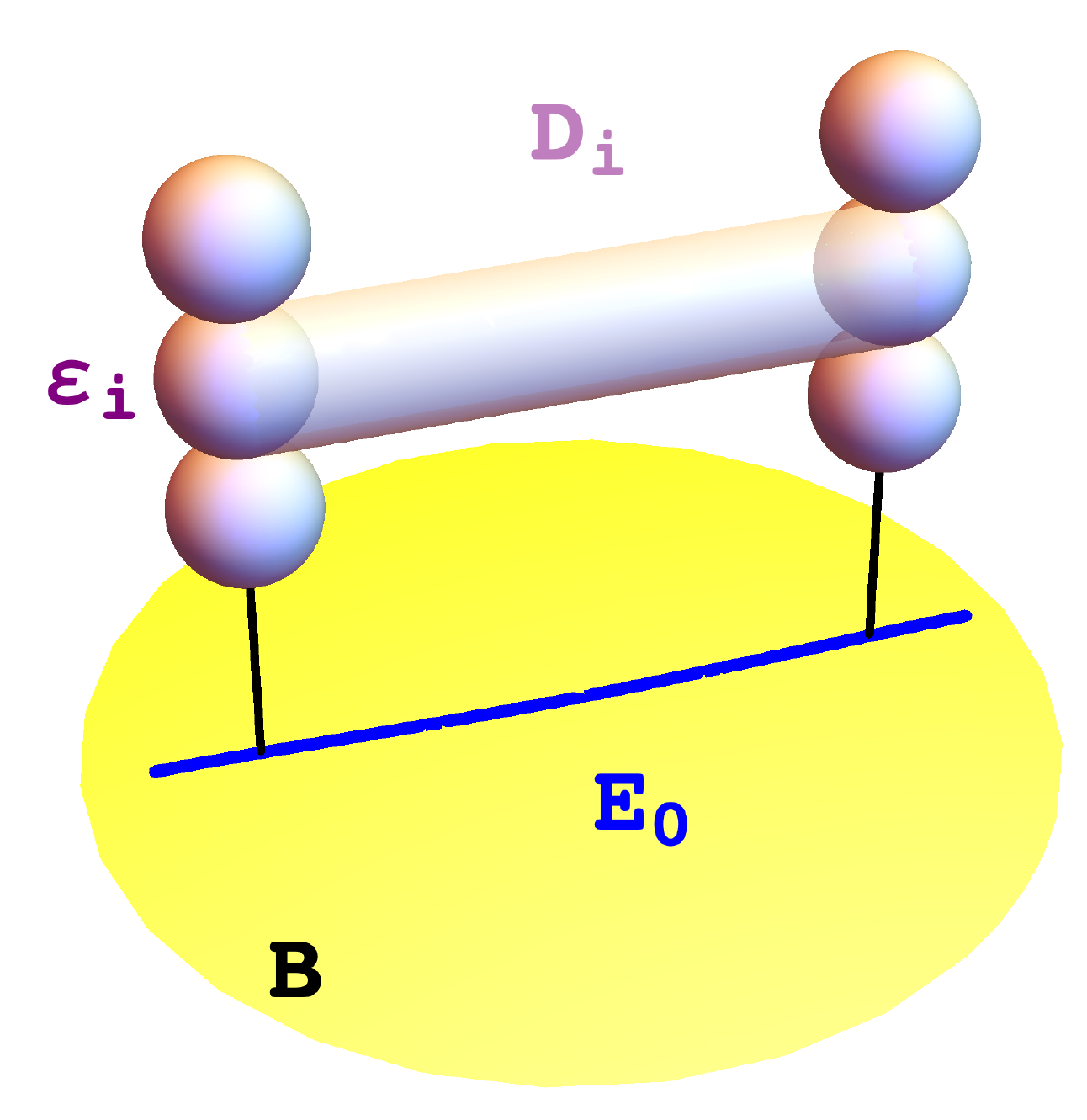}
\end{center}
\caption{Singular fibers of the I$_N^s$ elliptic Calabi-Yau threefolds. The base $B$ is an algebraic (compact or not) surface, and $E_0$ is the curve that supports the singular fiber. The singular fiber consists of $N$ copies of $\mathbb{P}^1$, denoted by $\varepsilon_i$, intersecting with each other as an affine $su(N)$ Dynkin diagram (in the figure we ignore the affine node). $D_i$ is the surface swept out by  $\varepsilon_i$ in the singular fiber along the curve $E_0$. The triple intersection numbers $D_i\cdot D_j\cdot D_k$ are the main quantities of interest in this paper.}\label{fig:eCY}
\end{figure}

The  low energy theory that arises from M-theory compactified on an I$_N^s$ model is a 5$d$ eight-supercharge theory with $su(N)$ gauge symmetry and matter fields. The vacuum moduli space of the theory has a Coulomb branch parametrized by the real scalars $\varphi$ in the  vector multiplet, which originates from the K\"ahler moduli of the internal Calabi-Yau threefold \cite{Cadavid:1995bk}. The Cartan gauge fields $A_i$ arise from decomposing the three-form in M-theory on the (1,1)-forms dual to $D_i$, where $D_i$ is the surface swept out by the $i$-th $\mathbb{P}^1$ in the I$_N^s$ singular fiber along the curve $E_0$ (see Figure \ref{fig:eCY}). Here $i=1,\cdots, N-1$ labels the Cartan of $su(N)$.

On a generic point on the Coulomb branch, the off-diagonal components of the gauge fields ($W$-bosons) and the matter fields are massive, and should be integrated out in the low energy effective action for the Cartan parts of the gauge fields. This Abelian effective action in particular contains the following 5$d$ Chern-Simons terms,\footnote{For 5$d$ theories coming from circle reductions of 6$d$ $\mathcal{N}=(1,0)$ theories, the 5$d$ gauge fields have three different kinds of 6$d$ origins. One is the Kaluza-Klein gauge field, while the others either come from the 6$d$ tensor fields or the 6$d$ gauge fields. We denote the 5$d$ gauge fields of the above three  kinds by $A_G, A_T, A_V$, respectively. The Chern-Simons terms that are of interest in the current paper are of the type $A_V\wedge F_V\wedge F_V$. The other types of Chern-Simons terms have been considered in \cite{Ferrara:1996wv,Bonetti:2011mw,Bonetti:2012fn,Bonetti:2013ela,Bonetti:2013cza,Grimm:2013oga}.}
\begin{align*}
{c_{ijk}(n_\mathbf{R}) \over 24\pi^2}\, A_i \wedge F_j \wedge F_k.
\end{align*}
 The Chern-Simons levels $c_{ijk}(n_\mathbf{R})$ are quantized and can be computed by standard one-loop Feynman diagrams \cite{Witten:1996qb,Intriligator:1997pq} (see also \cite{Bonetti:2013ela}).\footnote{Generally for 5$d$ theories coming from 6$d$,  the triangle diagrams with nonzero Kaluza-Klein modes running in the loop would contribute to the Chern-Simons levels. For example, the  Chern-Simons term $A_G\wedge F_G\wedge F_G$, where $A_G$ is the Kaluza-Klein gauge field, does receive such a contribution and is compared with the F/M-theory compactifications\cite{Bonetti:2013ela}. On the other hand, for the pure non-Abelian gauge field Chern-Simons terms $A_V\wedge F_V\wedge F_V$ we consider, the contributions from the nonzero Kaluza-Klein modes  cancel between the positive and negative Kaluza-Klein levels, so we will never have to worry about them.} 
In particular, $c_{ijk}(n_\mathbf{R})$ is a linear function of the multiplicities $n_\mathbf{R}$ of  hypermultiplets in each representation $\mathbf{R}$. The generator polynomial, known as the prepotential, for the Chern-Simons levels is given in \eqref{prepotential} and the $su(N)$ cases are listed in Table \ref{table:prepotential} for small $N$.

Geometrically, the Chern-Simons level $c_{ijk}(n_\mathbf{R})$ carries the interpretation as the triple intersection number  $D_i\cdot D_j\cdot D_k$ in the Calabi-Yau threefold \cite{Cadavid:1995bk,Witten:1996qb}. The main result of the current paper is an explicit calculation of the triple intersection numbers in the resolved I$_N^s$ elliptic Calabi-Yau threefolds. The key that enables this calculation is the recent results on  small resolutions of the I$_N^s$ Weierstrass models by sequences of blowups \cite{Esole:2011sm,Marsano:2011hv,Krause:2011xj,Morrison:2011mb,Tatar:2012tm,Lawrie:2012gg,Braun:2013cb,Hayashi:2013lra,Hayashi:2014kca,Esole:2014bka,Esole:2014hya,Braun:2014kla}. In particular, we will heavily rely on the resolutions given in \cite{Esole:2014bka,Esole:2014hya}. By repeatedly pushing forward \cite{fultonintersection,aluffi2010chern,fullwood2012stringy} the triple intersection via the blowup maps, one can express the intersection number $D_i\cdot D_j\cdot D_k$ solely in terms of the geometric data in the base, namely, the self-intersection  and the genus $g$ of the curve $E_0$. The final expression for the triple intersection number is presented in \eqref{finalint},
\begin{align*}
\begin{split}
D_i\cdot D_j \cdot D_k \,(E_0^2,g)=
\alpha_{ijk}\, E_0^2 +  \beta_{ijk} \,(2-2g),
\end{split}
\end{align*}
for some integers $\alpha_{ijk}$ and $\beta_{ijk}$ determined in \S\ref{app:intersection}. Here we have kept the  dependence of the triple intersection number $D_i\cdot D_j\cdot D_k$ on the geometric data in the base explicit. We list the  answers for small $N$ in Table \ref{table:intersection}.

Before continuing the discussion, we would like to point out  an interesting phenomenon under flop transitions. On the gauge theory side, the Coulomb branch is partitioned into different subchambers by real codimension one walls, where some matter fields become massless. Accordingly, the Chern-Simons levels $c_{ijk}(n_\mathbf{R})$ jump in going from one subchamber to another \cite{Witten:1996qb,Intriligator:1997pq}. On the geometry side, each subchamber is the relative K\"ahler cone of a resolved elliptic Calabi-Yau, and the codimension one locus is the junction between the relative K\"ahler cones of two different resolutions of the same singular Calabi-Yau, which are related by a flop transition. The discontinuities in the Chern-Simons levels $c_{ijk}(n_\mathbf{R})$ are precisely captured by the jumps in the triple intersection numbers $D_i\cdot D_j\cdot D_k$ under the flop transition. We demonstrate this phenomenon explicitly for the two resolved I$_3^s$ Weierstrass models in \S\ref{sec:chmatter}. In other parts of the paper, we focus on one particular resolution of the I$_N^s$ Weierstrass model (defined in \S\ref{sec:INmodel}) and on one particular subchamber on the Coulomb branch (defined in \eqref{SUNphase}).

Matching the Chern-Simons level with the triple intersection number,
\begin{align}\label{match}
c_{ijk} (n_\mathbf{R}) = D_i \cdot D_j\cdot D_k \,(E_0^2,g),
\end{align}
we note that  the left-hand side   is a function of the multiplicities $n_\mathbf{R}$ of the charged matter fields, while  the right-hand side  is a function of the geometric data in the base $B$.  The above equality then determines the multiplicities $n_\mathbf{R}$ of the charged matter fields  in the low energy 5$d$ theory obtained from M-theory compactified on the I$_N^s$ elliptic Calabi-Yau to be (see \eqref{mattercontent})\footnote{Here we assume $N\ge4$. The cases of $su(2)$ and $su(3)$ are special and presented in \eqref{su2matter}  and \eqref{su3matter}, respectively.}
\begin{align}\label{intromatter}
\begin{split}
&n_F = 16-16g +(8-N) E_0^2, ~~~~n_A= 2-2g+E_0^2,~~~~n_{\text{adj}}=g,
\end{split}
\end{align}
where $F$ and $A$ stand for the fundamental and antisymmetric representation, respectively.

Having obtained the multiplicities of the 5$d$ charged matter fields, we continue to discuss their potential 6$d$ origins. Generally a consistent 5$d$ eight-supercharge field theory cannot be lifted to a consistent 6$d$ $\mathcal{N}=(1,0)$ field theory. The obstructions are the gauge and gauge-gravitational mixed anomalies in 6$d$.\footnote{Except for \S\ref{sec:gravanomaly}, we  consider theories that may not be able to couple to gravity, therefore the cancellation of the pure gravitational anomalies is not essential. On the geometry side, we do not assume the compactness of the elliptic Calabi-Yau in the calculation of the triple intersection.} However, as we  show in \S\ref{sec:anomaly},  the 5$d$ $su(N)$ theories \eqref{intromatter} from M-theory on \textit{elliptic} Calabi-Yau threefolds can be lifted to anomaly-free 6$d$ theories.  Assuming that these 6$d$ parent theories come from compactifications on the same elliptic Calabi-Yau threefolds, the above results envision a 12$d$ origin, which as we know is F-theory. Similar ideas that explore the relation between the intersection numbers and anomaly cancellation have been considered in \cite{Park:2011ji,Bonetti:2011mw,Grimm:2011fx,Bonetti:2012fn,Bonetti:2013ela,Grimm:2013oga,Bonetti:2013cza,Grimm:2015zea}.

\begin{figure}[h!]
\begin{center}
\includegraphics[width=1\textwidth]{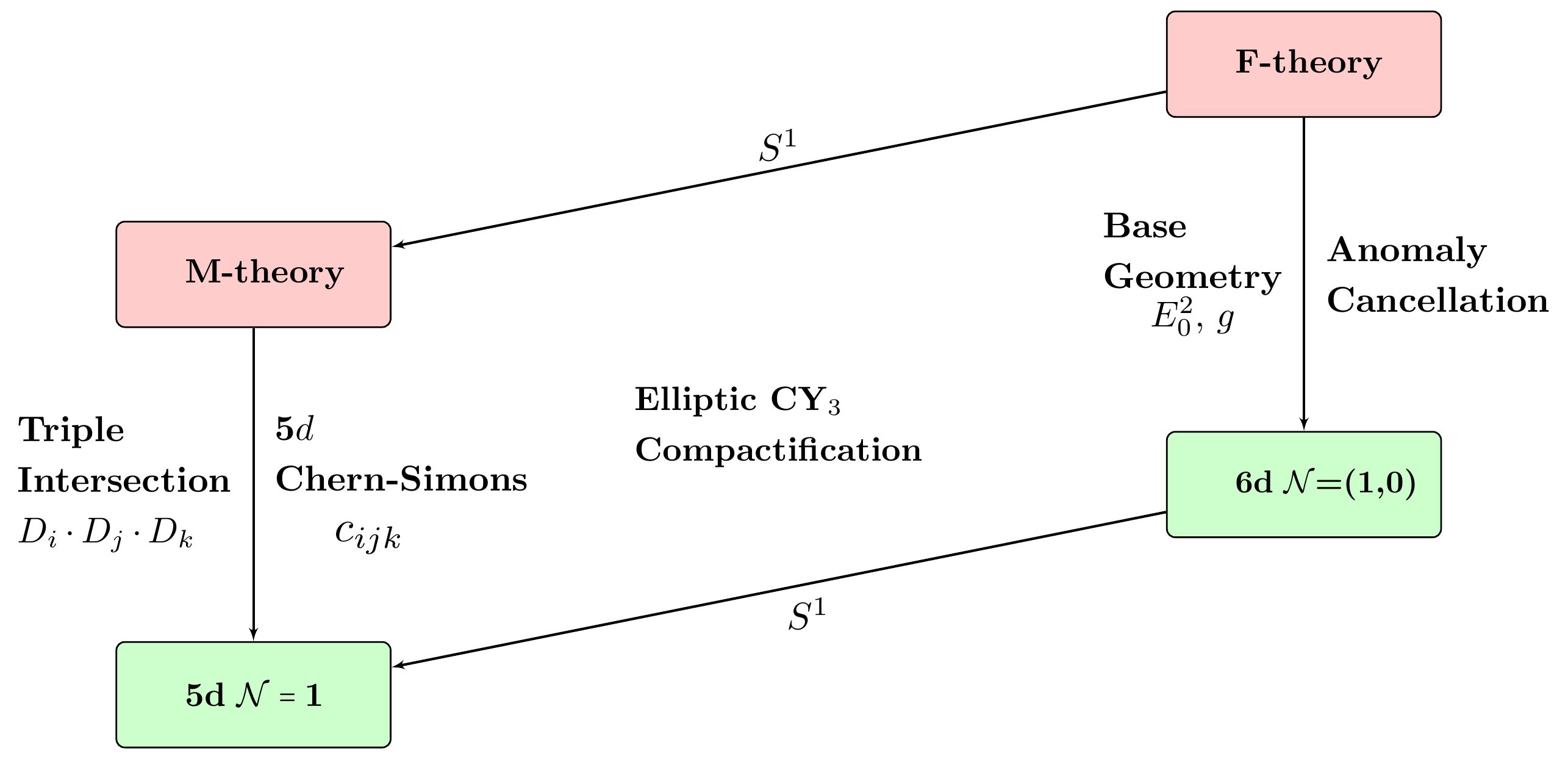}
\end{center}
\caption{M-theory and F-theory on elliptic Calabi-Yau threefolds. We determine the 5$d$ charged matter contents by matching the triple intersection numbers in the internal Calabi-Yau threefold with the 5$d$ Chern-Simons levels \eqref{match}.  On the other hand, the 6$d$ matter contents can be determined from anomaly cancellation equations. We show that the above diagram commutes for the I$_N^s$  elliptic Calabi-Yau threefolds with any algebraic base $B$.}\label{fig:flowchart}
\end{figure}

Our techniques in computing the triple intersection, summarized in \S\ref{app:intersection} , have a broad application to  other topological invariants and more general elliptic Calabi-Yau varieties. As an example, we compute the Euler characteristic of the resolved I$_N^s$ elliptic Calabi-Yau threefold. The answer agrees with a formula proved in \cite{grassi2000group,Grassi:2011hq} by explicitly computing the  contributions to the Euler characteristic from all singular fibers.\footnote{On the other hand, our calculation is based on a direct evaluation of the degree of the  top Chern class in the spirit of the Poincare-Hopf theorem. This has the advantage of avoiding a detailed analysis of the fibration structure.} It would be extremely exciting to extend our method to other physically interesting elliptic Calabi-Yau threefolds or fourfolds, for example in the contexts of non-Higgsable clusters \cite{Morrison:2012np,Morrison:2012js,Grassi:2014zxa,Morrison:2014lca} or F-theory constructions of 6$d$ superconformal field theories \cite{Heckman:2013pva,DelZotto:2014hpa,DelZotto:2014fia,Heckman:2014qba,Heckman:2015bfa}.

In the I$_N^s$ Weierstrass model, we have shown that the triple intersection numbers can be predicted by combining the anomaly cancellation equations and the field theory result  \eqref{prepotential} for the 5$d$ Chern-Simons levels. Assuming the F/M-theory duality is at work, we discuss a simple algorithm in \S\ref{sec:outlook} for a closed form expression of the triple intersection numbers in  elliptic Calabi-Yau threefolds. \\

The remaining of the paper is organized as follows. In \S\ref{sec:5d}, we review  the 5$d$ eight-supercharge gauge theories and their Coulomb branch effective actions.

  \S\ref{sec:Weierstrass} contains the main results of the paper. In \S\ref{sec:weierstrass} and \S\ref{sec:resolution}, we review the definition of the Weierstrass model and its resolutions.  In \S\ref{sec:intersection} and \S\ref{sec:euler}, we present the final expressions on the triple intersection numbers and Euler characteristics of the resolved I$_N^s$ Weierstrass model. 

In \S\ref{sec:Mtheory}, we discuss general aspects of M-theory compactification and determine the multiplicites $n_\mathbf{R}$ of  charged hypermultiplets in the 5$d$ low energy theory.  In \S\ref{sec:anomaly}, we show that the 5$d$ theories from M-theory compactified on I$_N^s$ elliptic Calabi-Yau threefolds can be lifted to anomaly-free 6$d$ theories.  In \S\ref{sec:outlook}, we exploit the F/M-theory duality to develop an algorithm in computing the triple intersection numbers in elliptic Calabi-Yau threefolds. In \S\ref{app:intersection} and \S\ref{app:Euler} we present the details of our calculation on the triple intersection and the Euler characteristic.

\section{5$d$ Coulomb Branch Effective Action}\label{sec:5d}

We will consider 5$d$ gauge theories with eight supercharges, focusing on the case with a vector multiplet\footnote{A word on conventions: we will refer to ``a vector multiplet with gauge group $G$" as $\text{dim }G$ vector multiplets that  transform in the adjoint representation of $G$. Similarly, ``a hypermultiplet in the representation $\mathbf{R}$" will stand for $\text{dim }\mathbf{R}$ hypermultiplets that transform in the representation $\mathbf{R}$ under the action of $G$.} with gauge group $G$  and $n_\mathbf{R}$ hypermultiplets in the representation $\mathbf{R}$. The  theory could be coupled to gravity or not, but we will only focus on its charged sector (with an exception in \S\ref{sec:gravanomaly}). As common in the world of eight supersymmetries, the vacuum moduli space has different branches. These include the Coulomb branch and the Higgs branch, parametrized by  scalars in the vector multiplet and the hypermultiplets, respectively. In addition to these two branches, there are also mixed Coulomb-Higgs branches where some of the scalars in both the vector multiplet and the hypermultiplets have nonzero vevs. We will focus on the Coulomb branch in this paper.

We denote the Lie algebra and the Cartan subalgebra of the gauge group by $\mathfrak{g}$ and $\mathfrak{h}\cong \mathbb{R}^r$, respectively, where $r$ is the rank of $G$. 
The real scalar in the vector multiplet will be denoted by $\varphi$, whose vev parametrizes the Coulomb branch. 

 At a generic point on the Coulomb branch, the gauge symmetry is broken by the vev of $\varphi$ to  $u(1)^r$, but there can be enhanced gauge symmetries at various special loci. After modding out the residual gauge symmetry, we can take $\varphi$ to be in the Weyl chamber $ \mathfrak{h}/W_G$ defined as
 \begin{align}
 \mathfrak{h}/W_G= \Big\{ \, \varphi\in \mathfrak{h}\, \Big\vert\, \varphi(\alpha)>0 ~~\text{for all positive roots } \alpha \,\Big\},
 \end{align}
 where $\varphi(\alpha)$ is the natural evaluation of $\varphi$ on the positive root $\alpha$.
 
At a generic point on the Coulomb branch, a hypermultiplet scalar $Q$ obtains a mass
\begin{align}
m_w= \varphi(w)
\end{align}
 from the vev of the real scalar $\varphi$.  Here $w$ is the weight in the representation $\mathbf{R}$ of the hypermultiplet scalar $Q$. We  integrate out these massive hypermultiplets as well as the off-diagonal parts of the vector multiplets ($W$-bosons) to obtain a Coulomb branch effective action for the Cartan vector multiplets.

 \subsection{Prepotentials and 5$d$ Chern-Simons Terms}

For 5$d$ gauge theories with eight supercharges, the metric $t(\varphi)_{ij}$ on the Coulomb branch is completely determined by a single real function called the \textit{prepotential} $\mathcal{F}(\varphi)$:
\begin{align}
t(\varphi)_{ij} = {\partial^2\mathcal{F}(\varphi)\over \partial \varphi_i \partial \varphi_j},
\end{align}
where $\varphi_i$, $i=1,\cdots, r$ is a basis for the Weyl chamber $\mathfrak{h}/W_G$. We will be more specific about the choice of the basis in a moment. The eight supersymmetries constrain the prepotential to be at most cubic in $\varphi$ \cite{Seiberg:1996bd}. The cubic term,
\begin{align}
{1\over 6}c_{ijk} \varphi_i\varphi_j\varphi_k,
\end{align}
in the prepotential $\mathcal{F}$ leads to an effective gauge coupling proportional to 
\begin{align}
t(\varphi)_{ij} \,F_i \wedge \star F_j,
\end{align}
and a Chern-Simons term
\begin{align}\label{CS}
{c_{ijk}\over 24\pi^2} A_i \wedge F_j\wedge F_k,
\end{align}
where $F_i$ is the field strength two-form of the $i$-th   Cartan gauge field $A_i$. 
The cubic coefficients are subject to the quantization condition \cite{Intriligator:1997pq}:
\begin{align}
c_{ijk}\in \mathbb{Z}.
\end{align}
The cubic order terms in the prepotential has two contributions. The first one is the classical (bare) Chern-Simons coupling
\begin{align}
{c_{cl}\over 6} d_{ijk} \varphi_i\varphi_j\varphi_k,
\end{align}
where $d_{ijk}={1\over 2} \Tr \left[T_i (T_jT_k +T_kT_j) \right]$ is the third-order Casimir of $\mathfrak{g}$. For gauge algebra other than ${su}(N)$ with $N\ge3$, there is no nontrivial third-order Casimir, hence $c_{cl}=0$. The second contribution comes from integrating out the massive modes at one-loop orders. Combining the above two contributions, the prepotential is determined to be \cite{Intriligator:1997pq}
\begin{align}\label{prepotential}
\mathcal{F}(\varphi) = {1\over 2} m_0 h_{ij} \varphi_i \varphi_j +{c_{cl}\over 6} d_{ijk} \varphi_i\varphi_j\varphi_k
+{1\over 12} \left(
\sum_{\alpha\in G} |\varphi(\alpha)|^3 -\sum_\mathbf{R}n_\mathbf{R} \sum_{w\in \mathbf{R}} | \varphi(w)|^3\right),
\end{align}
where the sum in $\alpha$ is over the roots of $G$ and the sum in $w$ is over the weights of $\mathbf{R}$. Here we have turned off the bare masses for the hypermultiplets. We will only focus on the cubic coefficients and ignore the quadratic term  in the following.

\subsection{Singularities on the Coulomb Branch}

There are two kinds of singularities on the Coulomb branch. The first kind is the boundary of the Coulomb branch $\varphi(\alpha)=0$ where some $W$-bosons become massless. The simplest example would be the Coulomb branch of an $su(2)$ gauge theory, which has the topology $\mathbb{R}/\mathbb{Z}_2$.\footnote{This is in sharp contrast to their 4$d$ children, the $su(2)$ Seiberg-Witten theories, whose Coulomb branch is a complex plane without boundary.} The second kind, which  is perhaps more interesting, is that the prepotential is not differentiable over various real codimension one \textit{walls} in the interior of the Coulomb branch  defined by\footnote{Note that since we take $\varphi$ to be in the Weyl chamber, the signs for $\varphi(\alpha)$ are always positive (negative) for positive (negative) roots in the interior of the Coulomb branch.}
\begin{align}\label{wall} 
\varphi(w)=0~~~\text{for}~w\in \mathbf{R}.
\end{align}
These codimension one walls \eqref{wall} are precisely the loci where some modes in the hypermultiplets become massless. At these loci we are no longer justified to integrate  them out  and the Coulomb branch metric becomes singular.

The real codimension one walls \eqref{wall} divide the Coulomb branch into several \textit{subchambers}. Given any gauge group $G$ and matter representations $\mathbf{R}$, we can then define the incident geometry constructed by the subchambers, the walls \eqref{wall}, and their intersections at higher codimensions \cite{Esole:2014bka,Esole:2014hya}.

\subsection{$su(N)$ Gauge Theories}\label{sec:sun}

In the following we focus on the $su(N)$ gauge theory with $n_F$ fundamental hypermultiplets, $n_A$ two-index antisymmetric hypermultiplets, and $n_{\text{adj}}$  adjoint hypermultiplets.\footnote{These are the most natural representations one can obtain from M-theory compactified on the I$_N^s$ elliptic Calabi-Yau threefolds. The fundamental and antisymmetric hypermultiplets  arise from  the collision of singular fibers in the elliptic fibration \cite{Katz:1996xe}. See the end of \S\ref{sec:INmodel} for discussions. On the other hand, the number of adjoint hypermultiplets is the genus $g$ of the curve in the base that supports the singular fiber \cite{Witten:1996qb} (see also \cite{Katz:1996ht}).}

 Let $D_i$ be the simple coroots and $\varepsilon_i$ be the simple roots of ${su}(N)$ with $i=1,\cdots,N-1$. The natural evaluation of $D_j$ on $\varepsilon_k$ will be denoted by $D_j(\varepsilon_k)$, which is nothing but the Cartan matrix
 \begin{align}
D_j( \varepsilon_k )= 
\begin{cases}
2~~~~~\text{if}~k=j,\\
-1~~~\text{if}~|k-j|=1,\\
0~~~~~\text{if}~|k-j|>1.
\end{cases}
\end{align}

We will use the simple coroots $D_i$ as our basis for the real scalar $\varphi$ of the vector multiplet:
\begin{align}\label{varphibasis}
\varphi = \sum_{i=1}^{N-1} \varphi_i D_i.
\end{align}
This is the natural basis to compare with the geometry side.

For $\varphi$ to be in the Weyl chamber $\mathfrak{h}/W_G$, we require $\varphi(\varepsilon_k)=  \sum_{i=1}^{N-1} \varphi_i D_i ( \varepsilon_k)>0$. The subchambers are most easily classified using another basis $a_k$ defined by
\begin{align}
a_k = \varphi_k-\varphi_{k-1},
\end{align}
with $\varphi_0=\varphi_N=0$. Note that $\sum_{k=1}^Na_k=0$. The statement that $\varphi_i$ lies in the Weyl chamber $\mathfrak{h}/W_G$ then translates into
\begin{align}
a_1>a_2>\cdots>a_N.
\end{align}

The prepotential for the $su(N)$ gauge theory  with $n_F$, $n_A$, $n_{\text{adj}}$ hypermultiplets in each representation can be  written in terms of $a_k$ as
\begin{align}\label{SUNpre}
\mathcal{F} (\varphi)= {1\over 6} \left[
(1-n_{\text{adj}}) \sum_{\substack{ i<j}}^N (a_i -a_j)^3 + c_{cl} \sum_{i=1}^N a_i^3 
-{ n_F\over2}\sum_{i=1}^N |a_i|^3- {n_A\over2} \sum_{i<j}^N |a_i+a_j|^3 
\right].
\end{align}
Each subchamber on the Coulomb branch is characterized by a particular choice of signs for the absolute values.

In the following we will write down the prepotentials for the $su(2)$ and $su(3)$ theories explicitly in each subchamber of the Coulomb branches. We will present our answer in the $\varphi_i$-basis \eqref{varphibasis}, which is the more natural basis when compared with the triple intersection numbers in the Calabi-Yau threefold. For the general $su(N)$ theory, we will present the prepotential only in one particular subchamber of the Coulomb branch.

\subsection*{$\bullet ~su(2)$}

For the $su(2)$ gauge theory with $n_F$ fundamental hypermultiplets and $n_{\text{adj}}$ adjoint hypermultiplets, the prepotential is \cite{Seiberg:1996bd,Douglas:1996xp,Morrison:1996xf}
\begin{align}\label{E1}
6\mathcal{F}(\varphi) = (8-8n_{\text{adj}}-n_F)\varphi_1^3,
\end{align}
where $\varphi_1$ is positive because we assume $\varphi$ to be in the Weyl chamber. 
The Coulomb branch has the topology of a half-line $\mathbb{R}/\mathbb{Z}_2$.

\subsection*{$\bullet ~su(3)$}

For the $su(3)$ gauge theory with $n_F$ fundamental hypermultiplets and $n_{\text{adj}}$ adjoint hypermultiplets, there are two subchambers on the Coulomb branch. We will denote these two subchambers by $\mathscr{T}$ and $\mathscr{T}'$, using the same symbols for their corresponding resolutions on the geometry side. They are defined by:
\begin{align}
&\mathscr{T}: ~\varphi_2>\varphi_1>{\varphi_2\over2}>0,\\
&\mathscr{T}': ~\varphi_1>\varphi_2> {\varphi_1\over2}>0,
\end{align}
or equivalently
\begin{align}\label{SU3phase}
&\mathscr{T}: ~a_1>a_2>0>a_3,\\
&\mathscr{T}': ~a_1>0>a_2>a_3.
\end{align}
 The prepotential  in the subchamber $\mathscr{T}$ is
\begin{align}\label{l=2}
\begin{split}
\mathscr{T}:~6\mathcal{F}(\varphi) &=( 8-8n_{\text{adj}}) \varphi_1^3 + (8-8n_{\text{adj}}-n_F) \varphi_2^3 + 3 \left( -1 +n_{\text{adj}}+ c_{cl} - {n_F\over 2} \right) \varphi_1^2 \varphi_2\\
& + 3\left(-1+n_{\text{adj}} - c_{cl} +{n_F\over2}\right) \varphi_1\varphi_2^2.
\end{split}
\end{align} 
 The prepotential  in the subchamber $\mathscr{T}'$ is
 \begin{align}\label{l=1}
 \begin{split}
\mathscr{T}':~6\mathcal{F}(\varphi)& = (8-n_{\text{adj}}+n_F) \varphi_1^3 +( 8-8n_{\text{adj}}) \varphi_2^3 
+ 3 \left( -1+n_{\text{adj}} + c_{cl} + {n_F\over 2} \right) \varphi_1^2 \varphi_2 \\
&
+ 3\left(-1+n_{\text{adj}} - c_{cl} -{n_F\over2}\right) \varphi_1\varphi_2^2.
\end{split}
\end{align} 
The coefficients are to be compared with the triple intersection numbers in the $su(3)$ resolutions $\mathscr{T}$ and $\mathscr{T}'$. The two subchambers are related by charge conjugation, which acts as $\varphi_1\rightarrow \varphi_2,~\varphi_2\rightarrow \varphi_1,~c_{cl}\rightarrow -c_{cl}.$ Hence it suffices to focus on one subchamber.

\subsection*{$\bullet ~su(N)$}

For the general $su(N)$ gauge theory with $n_F$ fundamental, $n_A$ antisymmetric, and $n_{\text{adj}}$ adjoint hypermultiplets, there are many different subchambers on the Coulomb branch. For now we will focus on one particular subchamber $\mathscr{T}$ defined by
\begin{align}\label{SUNphase}
\begin{split}
\mathscr{T}:~&a_1>\cdots a_{\,\left \lceil {N\over 2}\right\rceil} >0 >a_{\,\left\lceil {N\over 2}\right\rceil +1} >\cdots > a_N,\\
&a_i+a_j>0,~~~~\text{if}~i+j\le N+1~~\text{and} ~~j\neq N,\\
&a_i+a_j<0,~~~~\text{otherwise},
\end{split}
\end{align}
for every pair of integers $i<j$ with $i,j=1,\cdots, N$. In the $su(2)$ case, it is the unique chamber of the Coulomb branch. 
In the $su(3)$ case, it is the subchamber $\mathscr{T}$ defined in \eqref{SU3phase}.  This subchamber on the Coulomb branch will correspond to a particular resolution defined in \S\ref{sec:INmodel}.

\begin{spacing}{.999}

With the choice of signs in \eqref{SUNphase}, the prepotential  \eqref{SUNpre} becomes
\begin{align}\label{Tpre}
\begin{split}
\mathscr{T}:~~\mathcal{F} (\varphi)=& {1\over 6} \left[\,
(1-n_{\text{adj}}) \sum_{\substack{ i<j}}^N (a_i -a_j)^3 + c_{cl} \sum_{i=1}^N a_i^3
  \, - {n_F\over2}\sum_{i=1}^{\left\lceil {N\over2}\right\rceil} a_i^3
  \,+  {n_F\over2}\sum_{i=\left\lceil {N\over2}\right\rceil+1}^{N} a_i^3
\right.\\
&\left.
 - {n_A\over2} \sum_{\substack{i<j\\ i+j\le N+1}}^{N-1} (a_i+a_j)^3
  \,+ {n_A\over2} \sum_{\substack{i<j\\ i+j> N+1}}^{N-1} (a_i+a_j)^3
\,+  {n_A\over2} \sum_{i=1}^{N-1} (a_i+a_N)^3
\,\right],
\end{split}
\end{align}
where we recall that $a_k = \varphi_k - \varphi_{k-1}$ with $\varphi_0=\varphi_N=0$.

\end{spacing}

\section{Weierstrass Models and Resolutions}\label{sec:Weierstrass}

In \S\ref{sec:weierstrass} we review the definition of the Weierstrass model, with a particular focus on the description as a hypersurface in a projective bundle. In the case when the total space of the elliptic fibration is Calabi-Yau, it can be used to engineer a specific class of 5$d$  gauge theories with eight supercharges from M-theory compactification. 

In \S\ref{sec:resolution}, we review the work in \cite{Esole:2014bka, Esole:2014hya} on small resolutions of the I$_N^s$ Weierstrass model for small $N$. In \S\ref{sec:INmodel}, we consider an explicit small resolution  of the I$_N^s$ Weierstrass model  for arbitrary $N$ that preserves the flatness of the elliptic fibration when the base is a surface. This particular resolution is dual to the subchamber \eqref{SUNphase} on the Coulomb branch of the 5$d$ gauge theory.

In \S\ref{sec:intersection} and \S\ref{sec:euler}, we present our results on the triple intersection numbers and the Euler characteristics  in the resolved I$_N^s$ Weierstrass model. 

\subsection{Weierstrass Models}\label{sec:weierstrass}

A Weierstrass model $\mathscr{E}_0$ is an explicit presentation of an  elliptic fibration which admits a global section \cite{nakayama1987weierstrass}. Throughout this paper, we will take the base variety $B$ to be a  algebraic variety of complex dimension two.  The base $B$ can either be compact or non-compact.

Locally over  each point on the base $B$, the fiber is an elliptic curve defined by a plane cubic algebraic curve,
\begin{align}
\mathscr{E}_0:~ y^2z + a_1 xyz + a_3 yz^2 -( x^3+a_2 x^2z+a_4 xz^2 +a_6z^3) = 0,\label{tate}
\end{align}
where $[x:y:z]$ are the  homogeneous coordinates  of $\mathbb{P}^2$ parametrizing the fiber,  and the coefficients $a_i$ functions on the base $B$.   The Weierstrass model has an obvious global section given by $x=z=0$.

Globally, a Weierstrass model over $B$ requires a choice of a line bundle  $\mathscr{L}\rightarrow B$. Let $\mathscr{O}_B$ be the trivial line bundle over $B$. Define the vector bundle $V$ to be
\begin{align}
V=\mathscr{O}_B \oplus \mathscr{L}^2 \oplus \mathscr{L}^3 \rightarrow B.
\end{align}
Next, consider the projectivization of $V$ by replacing the fibers by  projective spaces,
\begin{align}\label{projectivebundle}
\pi:~\mathbb{P}( \mathscr{O}_B\oplus \mathscr{L}^2 \oplus \mathscr{L}^3)\rightarrow B.
\end{align}
We denote by $\mathscr{O}(1)\rightarrow \mathbb{P}(V)$  the canonical line bundle over the the projective bundle $\mathbb{P}(V)$.

In the global description above, the homogeneous coordinates $x,y,z$ and the coefficients $a_i$ are taken to be sections of the following line bundles over $\mathbb{P}(V)$:
 $$\left\{
\begin{tabular}{l}
 $z$ is a section of $\mathscr{O}(1)$,\\
$x$ is  a section of $\mathscr{O}(1)\otimes\pi^*\mathscr{L}^2$,\\
 $y$ is a section of $\mathscr{O}(1)\otimes\pi^* \mathscr{L}^3$,\\
 $a_i$ is a section of $\pi^* \mathscr{L}^i$.
\end{tabular}
\right.
$$ 
The Weierstrass model $\mathscr{E}_0$ is then described as the zero of the section given by  \eqref{tate} in the following line bundle over $\mathbb{P}(V)$:
 \begin{align}
\mathscr{O}(3) \otimes \pi^* \mathscr{L}^6.
\end{align}
This highbrow global description of the Weierstrass model in terms of a projective bundle will prove to be powerful in \S\ref{app:intersection} and \S\ref{app:Euler} when we compute the triple intersection numbers and the Euler characteristics. For a physics application, see \cite{Sethi:1996es}  for an example.

 \subsubsection*{Calabi-Yau Condition}

So far we have defined the Weierstrass model for a general line bundle $\mathscr{L}$ over the base $B$. We would now like to impose the Calabi-Yau condition on the Weierstrass model to fix $\mathscr{L}$.

Let $H$ be the divisor class for $\mathscr{O}(1)$ over $\mathbb{P}(V)$, and $L$ be the divisor class of the line bundle $\mathscr{L}\rightarrow B$, \textit{i.e.} $L:=c_1(\mathscr{L})$. The total Chern classs of the the Weierstrass  model $\mathscr{E}_0$ can be obtained by applying the adjunction formula,
\begin{align}
c(\mathscr{E}_0)  ={ (1+H) (1+H+2\pi^* L)(1+H+3\pi^* L)  \over (1+3H +6\pi^*L) } \pi^*c(B).
\end{align}
In particular, the first Chern class of $\mathscr{E}_0$ is
\begin{align}
c_1(\mathscr{E}_0)  = \pi^*(-K-L),
\end{align}
where $-K$ is the anticanonical class of the base $B$. 
 Hence the Weierstrass  model $\mathscr{E}_0$ is Calabi-Yau only when 
\begin{align}
L=-K,
\end{align}
that is, when $\mathscr{L}$ is the anticanonical line bundle of $B$.

\subsubsection*{Singular Fibers and Tate Forms}

Given a Weierstrass model \eqref{tate},  the discriminant $\Delta$ and the $j$-invariant are defined as
\begin{align}
\Delta &= -b_2^2 b_8 -8 b_4^3 -27 b_6^2 + 9 b_2 b_4 b_6=\frac{1}{1728} (c_4^3-c_6^2), \\
  j &=\frac{c_4^3}{\Delta},
\end{align}
where $(b_2,b_4,b_6)$ or $(c_4,c_6)$ are defined in terms of the sections $a_i$ in \eqref{tate},
\begin{align}\label{eq.formulaire}
\begin{split}
b_2 &= a_1^2+ 4 a_2,~~b_4 = a_1 a_3 + 2 a_4 ,~~~b_6  = a_3^2 + 4 a_6 , ~~~b_8  =b_2 a_6 -a_1 a_3 a_4 + a_2 a_3^2-a_4^2,\\
c_4 &= b_2^2~ -24 b_4, ~~~c_6 = -b_2^3+ 36 b_2 b_4 -216 b_6,\\
\end{split}
\end{align}
The fibration of the Weierstrass model is singular over the discriminant locus  $\Delta=0$. 

 A nonsingular  Weierstrass model only has nodal and cuspidal curves as  singular fibers. In order to have more interesting singular fibers, we have to consider singular (as a total space)  Weierstrass models. 
The singularity of an elliptic fibration over codimension one loci (\textit{i.e.} divisors) of the base are classified by Kodaira \cite{kodaira1963II,kodaira1963III} and N\'eron \cite{neron} and determined by Tate's algorithm \cite{tate1975algorithm}. Specifically , we can enforce a given singularity over a curve $E_0$ in $B$: 
\begin{align}
E_0:~e_0=0
\end{align}
by allowing the coefficients $a_i$ to vanish on $E_0$ with certain multiplicities.  Given the multiplicities of $e_0$ for each of the sections $a_i$, the type of singularity is determined by Tate's algorithm. If $a_i$ has multiplicity $k$, we will write
\begin{align}
a_i = a_{i, k } e_0^k.
\end{align}
In the case $k=0$ we will simply write $a_{i,k}$ as $a_i$.

In this paper we will consider  Weierstrass models with singular fibers of the type I$_N^s$, which corresponds to the gauge group $su(N)$. 
The multiplicities of $a_i$ for type I$_{2n}^s$  and type I$_{2n+1}^s$ models are \cite{Bershadsky:1996nh,Katz:2011qp}:
\begin{align}
\text{I}_{2n}^s:~&a_1 = a_1 ,~a_2 = a_{2,1} e_0,~a_{3} = a_{3,n }e_0^n,~a_4 = a_{4,n}e_0^n,~a_6 = a_{6,2n} e_0^{2n},\\
\text{I}_{2n+1}^s:~&a_1 = a_1 ,~a_2 = a_{2,1} e_0,~a_{3} = a_{3,n }e_0^n,~a_4 = a_{4,n+1}e_0^{n+1},~a_6 = a_{6,2n+1} e_0^{2n+1}.
\end{align}
After a resolution of singularities, the singular fiber of type I$_{N}^s$  consists of $N$ copies of $\mathbb{P}^1$ with intersection matrix being the affine Dyknin diagram of $su(N)$. We will consider an explicit resolution in \S\ref{sec:INmodel}.

\subsection{Resolutions}\label{sec:resolution}

Given a Weierstrass model with a singular fiber type, the total space is generally singular. To completely resolve the singularity, one needs to blow up the singularity repeatedly. At each step of blowups, there are generally more than one ways to proceed. All these different choices of blowups at each step then form a \textit{network of resolutions}.

In the resolved space, the  fibration is still singular over the curve $E_0$ while the total space is nonsingular. The singular fiber consists of $N$ copies of $\mathbb{P}^1$ intersecting in the way as an affine $su(N)$ Dynkin diagram. Let $D_i$ ($i=0,1,\cdots, N-1$) be the \textit{surfaces}
 swept out by the $\mathbb{P}^1$'s  in the singular fiber along the curve $E_0$.\footnote{As we will seen in \S\ref{sec:Mtheory}, the surface $D_i$ is identified as the simple coroot of ${su}(N)$ in M-theory compactification, hence we will use the same symbol $D_i$ for both the surface and the coroot. } 
 The divisor classes for $D_i$ in the threefold can be read off from the centers of the blowups. They are expressed in terms of  the exceptional divisor class $E_i$ for each  blowup and (pullback of) the   divisor class $E_0$ in the base $B$. We will present the explicit expression for $D_i$ in the following.

\subsubsection{Lower Rank Cases}

 The network of resolutions for the I$_N^s$ Weierstrass models with small $N$ are studied in \cite{Esole:2014bka, Esole:2014hya}. The cases for I$_2^s$ and I$_3^s$ models are shown in Figure \ref{tree2} and Figure \ref{tree3}. Each arrow in the figures represents a blowup. The variables  above the arrow but left to the bar are the generators for the center of the blowup. $E_i:~e_i=0$ is the exceptional divisor for the $i$-th blowup. For example, a blowup represented by $(x,y,e_0|e_1)$ is obtained by replacing the variables $x,y,e_0$ by projective coordinates $[\tilde x: \tilde y:\tilde e_0]$,
 \begin{align}
 x=e_1 \tilde x,~~~~y=e_1 \tilde y,~~~~ e_0 = e_1\tilde e_0.
 \end{align}
 For simplicity of notations, we will then drop the tilde for the projective coordinates after each step of blowup and forget the original coordinates.

\begin{figure}
\begin{center}
\begin{tikzcd}[column sep=3cm]
\displaystyle{ \mathscr{E}_0}\arrow[leftarrow]{r}{\displaystyle (x,y,e_0|e_1)} &  \mathscr{T},
\end{tikzcd}
\end{center}
\caption{The network of resolutions for the I$_2^s$ model. Each letter stands for a (partial) resolution and each arrow represents a blowup. Starting from $\mathscr{E}_0$, there is a unique (small)  resolution $\mathscr{T}$.}\label{tree2}

\begin{center}
\scalebox{1}{
\begin{tikzcd}[column sep=huge,  ampersand replacement=\&]
\& \& \mathscr{T}  \arrow[bend left=75, leftrightarrow, dashed, color=red]{dd}[right] {\quad \text{\large flop}} \\
\mathscr{E}_0\arrow[leftarrow]{r}{\displaystyle (x,y,e_0|e_1)} \& \mathscr{E}_1 \arrow[leftarrow,sloped, near end ]{ur}{\displaystyle(y,e_1|e_2)} \arrow[leftarrow, near start, sloped]{dr}{\displaystyle (s,e_1|e_2)} \& \\\
\&    \& \mathscr{T}'
\end{tikzcd}}
\end{center}
\caption{The network of resolutions for the I$_3^s$ model. Each letter stands for a (partial) resolution and each arrow represents a blowup. Starting from $\mathscr{E}_0$, there is a unique (crepant) blowup $(x,y,e_0|e_1)$ to go to the partial resolution $\mathscr{E}_1$. For the second step, there are two inequivalent blowups leading to $\mathscr{T}$ and $\mathscr{T}'$. The two resolutions $\mathscr{T}$ and $\mathscr{T}'$ are related by a flop induced by the $\mathbb{Z}_2$ automorphism in the Mordel-Weil group. Here $s=y+a_1x+a_{3,1}e_0$.}\label{tree3}
\end{figure}

The classes for the surfaces $D_i$ can be determined from the centers of blowups. For example, in the I$_2^s$ model, the class for the surface $D_0$ corresponds to the proper transform of $e_0=0$ in the original Weierstrass model $\mathscr{E}_0$ \cite{Esole:2014bka, Esole:2014hya}. In the resolved space $\mathscr{T}$, the divisor class for $D_0$ is then (the pullback of) $E_0$ with one factor of the exceptional divisor $E_1$ stripped off, \textit{i.e.} $D_0=E_0-E_1$. Similarly we can immediately read off the divisor classes for the other surfaces $D_i$ in the I$_2^s$ and I$_3^s$ models:
\begin{align}\label{Dclass}
\begin{split}
&su(2):\\
&~~~~~~\mathscr{T}:~\,\,\, D_0 = E_0-E_1,~~~D_1 = E_1,\\
&su(3):\\
&~~~~~~\mathscr{T}:~\,D_0=E_0-E_1,~~~D_1=E_1-E_2,~~~D_2=E_2,\\
&~~~~~~\mathscr{T}':~D_0=E_0-E_1,~~~D_1=E_2,~~~~~~\,~~~D_2=E_1-E_2,\\
\end{split}
\end{align}

\subsubsection{Resolution of the I$_N^s$ Model}\label{sec:INmodel}

For the  I$_N^s$ Weierstrass model with general $N$, there are many different small resolutions and the explicit constructions for every resolution would be quite tedious. Here we will consider one particular small resolution  $\mathscr{T}$ of the I$_N^s$ Weierstrass model that preserves the flatness\footnote{Over the field $\mathbb{C}$, a fibration is called \textit{flat} if the fiber is equidimensional.} of the elliptic fibration when the base is a surface. For the I$_{2n}^s$ fiber, \textit{i.e.} $N=2n$, the resolution $\mathscr{T}$ is defined by
\begin{align}\label{I2n}
\mathscr{T} : ~\mathscr{E}_0 \, \underset{f_1}{\xleftarrow{(x,y,e_0|e_1) } }
\, \mathscr{E}_1\,\underset{f_2}{ \xleftarrow{(y,e_1|e_2) }}
 \,\mathscr{E}_2 \,\underset{f_3}{\xleftarrow{(x,e_2|e_3)} }
  \cdots
 \,\underset{f_{2n-1}}{ \xleftarrow{(x,e_{2n-2}|e_{2n-1})}}\,
  \mathscr{T}
\end{align}
For the I$_{2n+1}^s$ fiber, \textit{i.e.} $N=2n+1$, the resolution $\mathscr{T}$ is defined by
\begin{align}\label{I2n1}
\mathscr{T} : ~\mathscr{E}_0 \,\underset{f_1}{\xleftarrow{(x,y,e_0|e_1) }}
\,\mathscr{E}_1\,\underset{f_2}{ \xleftarrow{(y,e_1|e_2) }}
 \,\mathscr{E}_2\, \underset{f_3}{\xleftarrow{(x,e_2|e_3)} }
\,\cdots \, \underset{f_{2n}}{\xleftarrow{(y,e_{2n-1}|e_{2n})}}
 \, \mathscr{T}
\end{align}
With the exception of the first blowup $f_1$, the center for the $2k$-th blowup is $(y,e_{2k-1})$ and for the $(2k+1)$-th blowup is $(x,e_{2k})$, where $e_i=0$ is the exceptional divisor. 
 Note that the number of blowups is $N-1$, which is the rank of $su(N)$. This particular resolution of I$_N^s$ model corresponds to the subchamber given in \eqref{SUNphase} of the 5$d$ Coulomb branch.

 In fact, the resolutions \eqref{I2n} and \eqref{I2n1} given above are isomorphic to those in \cite{Lawrie:2012gg}. Even though the centers of the blowups at each step are very different , the isomorphism can be shown by studying  the scalings of the variables with respect to the projective spaces introduced by the blowups.\footnote{The exceptional divisors in \cite{Lawrie:2012gg} are related to ours by $e_0 = \zeta_0,~e_{2i-1} =\zeta_i,~ e_{2j}=\delta_j$, with $i=1,\cdots, n$ and $j=1,\cdots,n-1$ for the  I$_{2n}^s$ model and $j=1,\cdots, n$ for the I$_{2n+1}^s$ model.} See \cite{Esole:2014bka,Esole:2014hya} for discussions and examples on isomorphisms between resolutions.

The resolved I$_{2n}^s$ model is a hypersurface defined by
\begin{align}\label{2nresolution}
\begin{split}
\mathscr{T}:~&y(y\prod_{i=1}^{n-1} e_{2i}  +a_1 x+ a_{3,n}e_0^n \prod_{i=1}^{n-1}e_{2i}^{n-i}\prod_{i=1}^{n-1} e^{n-i}_{2i-1}  )=
x^3\prod_{i=1}^n e_{2i-1}^{i} \prod_{i=2}^{n-1} e_{2i}^{i-1}\\
&
+a_{2,1} e_0x^2\prod_{i=1}^n e_{2i-1}  
+ a_{4,n}  e_0^nx \prod_{i=1}^{n-1} e^{n-i}_{2i-1} \prod_{i=1}^{n-2} e_{2i}^{n-i-1} 
+a_{6,2n} e_0^{2n} \prod_{i=1}^{n-1} e_{2i-1}^{2n-2i} \prod_{i=1}^{n-1} e_{2i}^{2n-2i-1},
\end{split}
\end{align}
in the blowup of the projective bundle $\mathbb{P}(V)$. Similarly, the resolved I$_{2n+1}^s$ model is a hypersurface defined by
\begin{align}\label{2n1resolution}
\begin{split}
\mathscr{T}:~&y\left(
y\prod_{i=1}^n e_{2i}  + a_1x + a_{3,n} e_0^n \prod_{i=1}^{n-1} e_{2i}^{n-i} \prod_{i=1}^{n-1} e_{2i-1}^{n-i} 
\right)
=x^3 \prod_{i=1}^{n} e_{2i-1}^{i} \prod_{i=2}^n e_{2i}^{i-1}\\
&
+a_{2,1}e_0x^2 \prod_{i=1}^{n} e_{2i-1}
+a_{4,n+1} e_0^{n+1} x\prod_{i=1}^n e_{2i-1}^{n-i+1} \prod_{i=1}^{n-1} e_{2i}^{n-i}
+a_{6,2n+1} e_0^{2n+1} \prod_{i=1}^{n}e_{2i-1}^{2n-2i+1}\prod_{i=1}^{n-1} e_{2i}^{2n - 2i},
\end{split}
\end{align}
in the blowup of the projective bundle $\mathbb{P}(V)$.

\begin{figure}[htb]
\centering
\subfloat[]{
\includegraphics[scale=1.25]{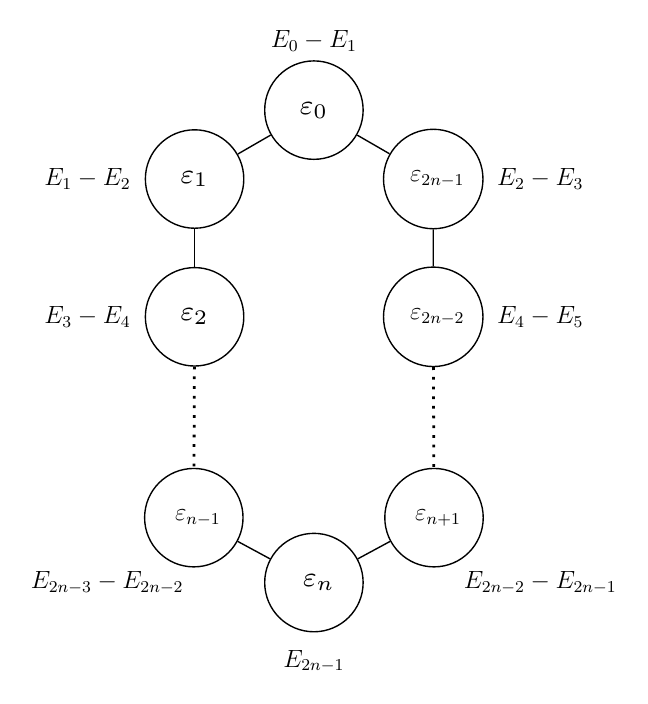}
}
~~~
\subfloat[]{
\includegraphics[scale=1.25]{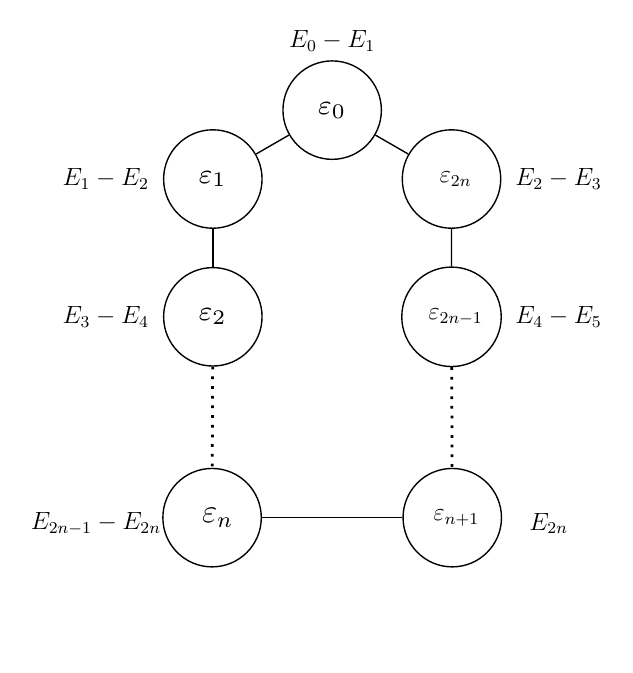}
}
\caption{(a) The singular fiber for the resolved I$_{2n}^s$ model $\mathscr{T}$ \eqref{I2n}. (b) The singular fiber for the resolved I$_{2n+1}^s$ model $\mathscr{T}$ \eqref{I2n1}. Here $\varepsilon_i$  is the fiber class of the surface $D_i$, each of which is a copy of $\mathbb{P}^1$. The classes  for $D_i$   are labeled next to the nodes (see \eqref{Cclass}). $E_i$ is the exceptional divisor for the $i$-th blowup, while the subscript of $\varepsilon_i$ labels  the position of the node in the affine $su(N)$ Dynkin diagram. }\label{fig:IN}
\end{figure}

As can be checked straightforwardly \cite{Lawrie:2012gg}, the resolved I$_N^s$ Weierstrass models considered above preserves the flat fibration for all $N$ if the total space is a threefold.\footnote{However, the resolved I$_N^s$ Weierstrass model we consider does not admit a flat fibration for $N\ge7$ if the total space is a fourfold. The fibration becomes non-flat over the codimension three locus $e_0=a_1=a_{2,1}=0$ in the base $B$. } The singular fibers over a generic point in $E_0$ for the resolved I$_N^s$ models  are shown in Figure \ref{fig:IN}. The classes for the surfaces $D_i$  swept out by the $i$-th node can be read off from the centers of the blowups:
\begin{align}\label{Cclass}
D_i = 
\begin{cases}
\,E_{2i-1} -E_{2i},~~\,~~~~~~~~~~\text{if}~~i<\left\lceil {N\over 2}\right\rceil ,\\
\, E_{N-1},~~~~~~~~~~~\,~~~~~~~~\text{if}~~i=\left\lceil {N\over 2}\right\rceil,\\
\,E_{2N-2i} -E_{2N-2i+1}.~\,~~~\text{if}~~i>\left\lceil {N\over 2}\right\rceil.
\end{cases}
\end{align}

\begin{figure}
\begin{center}
\scalebox{1}{
\begin{tikzcd}[column sep=huge,  ampersand replacement=\&]
 \&~~ \text{\bf I}_{N+1}^s  :~ e_0=P_N=0 \\
\text{\bf I}_N^s :~ e_0=0\arrow[rightarrow,sloped, near end ]{ur}{} \arrow[rightarrow, near start, sloped]{dr}{} \& \\\
   \&~~ \text{\bf I}^*_{N-4}:~ e_0=a_1=0\\
\text{codim 1}\&\text{codim 2}
\end{tikzcd}}
\end{center}
\caption{The singular fibers of the I$_N^s$ Weierstrass model. Over the codimension one locus $e_0=0$ in the base, the singular fiber is of the type I$_N^s$. Over the codimension two singular loci $e_0=a_1=0$ and $e_0=P_N=0$, the singular fiber enhances to I$_{N-4}^*$ and I$_{N+1}^s$, respectively.}\label{fig:codim2}
\end{figure}
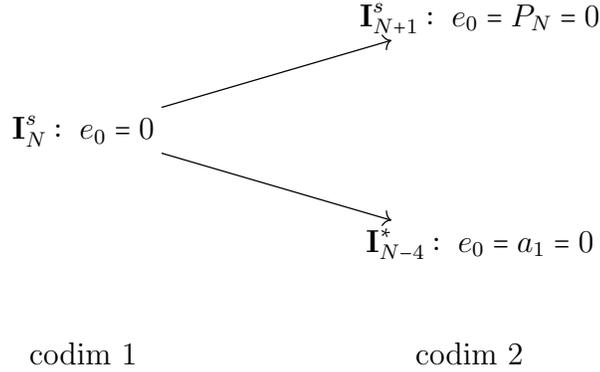

Let us take a close look at the discriminant locus in the type I$_N^s$ Weierstrass model. 
The  discriminant  factorizes into two components:
\begin{equation}
\Delta=e_0^{N} \Big[-a_1^4 P_{N}+\mathcal{O}(e_0)\Big],
\end{equation}
where
\begin{align}
P_N = \begin{cases}
-a_1a_{3,n} a_{4,n} -a_{4,n}^2 +a_1^2a_{6,2n},~~~\,~~~~~~~~~\text{if}~~N=2n,\\
a_{2,1} a_{3,n}^2 - a_1 a_{3,n} a_{4,n+1} + a_1^2 a_{6,2n+1},~~~~~\text{if}~~N=2n+1.
\end{cases}
\end{align}
 The first component $e_0^{N}$ is the codimension one locus over which we have the fiber of type I$_{N}^s$.  The second component $[-a_1^4 P_{N}+\cdots]$ is the codimension one locus over which we have the nodal curves I$_1$. 
These two components intersect in codimension two in the base along $e_0=a_1=0$ and $e_0=P_{N}=0$, over which one obtains enhanced singular fibers of types I$_{N-4}^*$ and   I$_{N+1}^s$, respectively.\footnote{The codimension two collision at $e_0=a_1=0$ is special for I$_2^s$ and I$_3^s$. The fiber enhancements in these two cases are I$_2^s\rightarrow$III and I$_3^s\rightarrow$IV, respectively \cite{Esole:2014bka}. } See Figure \ref{fig:codim2} for illustrations.

 These fiber enhancements can be seen straightforwardly from the resolutions  \eqref{2nresolution} and \eqref{2n1resolution}. As can be checked explicitly, the node $e_{N-1}=0$ is a conic with $P_N$ being its discriminant. Over the codimension two locus $e_0=P_N=0$, the conic $e_{N-1}=0$ splits into two nodes inducing the fiber enhancement I$_N^s\rightarrow$I$_{N+1}^s$. On the other hand, the curve $E_0$ intersects with the cuspidal locus $c_4=c_6=0$ at $e_0=a_1=0$, where we have the fiber enhancement I$_N^s\rightarrow$I$_{N-4}^*$. The lower rank examples can be found in details in \cite{Esole:2014bka,Esole:2014hya}.

Physically, the codimension two collision has the interpretation of gauge symmetry enhancement $su(N) \rightarrow so(2N)$ and $su(N)\rightarrow su(N+1)$ over these loci. These enhancements indicate extra matter field degrees of freedom, whose representations can be determined by decomposing the adjoint representation of the larger symmetry group into the original one \cite{Katz:1996xe}.  For example, the rank-one enhancements  $su(N) \rightarrow so(2N)$ and $su(N)\rightarrow su(N+1)$ in the I$_N^s$ elliptic Calabi-Yau give rise to matter fields in the two-index antisymmetric and fundamental representations, respectively \cite{Bershadsky:1996nh,grassi2000group}. This justifies our choice of the representations for the 5$d$ gauge theories in \S\ref{sec:sun}.

\subsection{Triple Intersection}\label{sec:intersection}

In this subsection we present the main result of the current paper, the triple intersection numbers of the surfaces $D_i$. We will describe the general strategy of the calculation and present the final result here. The details are left to \S\ref{app:intersection}.

The resolved Weierstrass model   is described by a hypersurface in the ambient fourfold $Y_{N-1}$ obtained by a sequence of blowups \cite{Esole:2014bka,Esole:2014hya}:
\begin{align}
\begin{split}
&Y_0 \xleftarrow{~f_1~} Y_1 \xleftarrow{~f_2~}Y_2\xleftarrow{~f_3~} \cdots \xleftarrow{~f_{N-1}~} Y_{N-1}\,\\
&\big\downarrow\,\pi\\
&B
\end{split}
\end{align}
Here $Y_0$ is the projective bundle\footnote{In the Calabi-Yau case, the line bundle  $\mathscr{L}$  is  the anticanonical line bundle of the base $B$. However, we will keep it general for now since the calculation for the triple intersection does not require the Calabi-Yau condition.} $\pi:\mathbb{P}(\mathscr{O}_B\oplus \mathscr{L}^2 \oplus \mathscr{L}^3)\rightarrow B$  and $f_i$ are the blowup maps. The singular Weierstrass model $\mathscr{E}_0$ is a hypersurface in $Y_0$.  

The general strategy to compute $D_i\cdot D_j\cdot D_k$ is  to pushforward the intersection numbers to the base $B$. The pushforward maps in the current case are induced by either the projection map $\pi$ or the blowup maps $f_{i}$. The triple intersection numbers at the end of the day can be written as
\begin{align}
\begin{split}
D_i\cdot D_j\cdot D_k&=\pi_* \circ f_{1*} \circ\cdots \circ f_{N-1*}(D_i\cdot D_j\cdot D_k \cap [\mathscr{T}] \,) \\
&= A_{ijk} \,  E_0\cdot E_0 + B_{ijk} \,  E_0 \cdot L,
\end{split}
\end{align}
for some integers $A_{ijk},B_{ijk}$. Here $[\mathscr{T}]$ is the class of the resolved I$_N^s$ model inside the ambient fourfold $Y_{N-1}$. In the Calabi-Yau case, $L$ is the anticanonical class of the base $B$, $L=-K$, and we have
\begin{align}
L \cdot E_0 = - K \cdot E_0 =E_0^2 +2 -2g,
\end{align}
where $g$ is the genus of the curve $E_0$. Hence in the Calabi-Yau case, the triple intersection numbers are integral linear combinations of $E_0^2$ and $g$:
\begin{align}
D_i \cdot D_j \cdot D_k =  \alpha_{ijk}\, E_0^2 +  \beta_{ijk} \,(2-2g) ,
\end{align}
for some integers $\alpha_{ijk}$ and $\beta_{ijk}$.
 The final expression for the triple intersection numbers are given by \eqref{finalint}. Even though we do not have a closed form formula for the triple intersection numbers in $\mathscr{T}$,  they can be computed straightforwardly for any given $N$.

The triple intersection numbers can be most conveniently encoded in  the \textit{triple intersection form} defined as:
\begin{align}
(\sum_{i=1}^{N-1} \varphi_i D_i)^3 = \sum_{i,j,k=1}^{N-1} D_i\cdot D_j\cdot D_k \, \varphi_i \varphi_j \varphi_k.
\end{align}
The coefficient of $\varphi_i \varphi_j \varphi_k$ in the triple intersection form is 6 times the triple intersection number $D_i\cdot D_j\cdot D_k$ if $i,j,k$ are distinct, 3 times $D_i\cdot D_j\cdot D_k$ if only two of $i,j,k$ are the same, and equal to $D_i\cdot D_j\cdot D_k$ if $i=j=k$.  We list the triple intersection number forms for the resolved I$_N^s$ model $\mathscr{T}$ for small $N$ in Table \ref{table:intersection}.  On the other hand, we list the prepotentials for the 5$d$ $su(N)$ gauge theory  with small $N$ in Table \ref{table:prepotential} for ease of comparison in \S\ref{sec:chmatter}.

\begin{table}[h!]
\begin{align*}
\left.\begin{array}{|c|c|}
\hline  \text{Singular Fiber} &\text{Triple Intersection Form}~ ~(\sum_i \varphi_i D_i)^3 \\
\hline \,\text{I}_2^s\, &  (-8+8g -6E_0^2)\varphi_1^3\\
\hline \,\text{I}_3^s\, &  (8-8g)\varphi_1^3+(-10+10g -6E_0^2)\varphi_2^3+3( -10+10g -3E_0^2)\varphi_1^2\varphi_2\\
&+3( 8-8g +3E_0^2)\varphi_1\varphi_2^2 \\
\hline\, \text{I}_4^s\, & (8-8g) \varphi_1^3 + (-8+8g-4E_0^2)\varphi_2^3 +(4-4g-2E_0^2)\varphi_3^3  \\
& +3(-6+6g -E_0^2 ) \varphi_1^2 \varphi_2
+3(4-4g +E_0^2)\varphi_1\varphi_2^2
+3( -4+4g -2E_0^2) \varphi_1^2\varphi_3\\
&+3(4-4g +E_0^2 ) \varphi_2^2 \varphi_3
+3(-6+6g -E_0^2)\varphi_2\varphi_3^2 + 6(2-2g+E_0^2)\varphi_1\varphi_2\varphi_3\\
\hline \,\text{I}_5^s\, & 
(8-8 g) \varphi_1^3
+\left(6-6g-E_0^2\right)\varphi_2^3 
+ \left(-8+8g-3 E_0^2\right)\varphi_3^3
+ \left(4-4g-2 E_0^2\right)\varphi_4^3\\
&+3  \left(4-4g+E_0^2\right) \varphi_1\varphi_2^2
+3\left(-6+6g-E_0^2\right)  \varphi_1^2 \varphi_2 
+3 \left(-2+2g-E_0^2\right)\varphi_2 \varphi_4^2 
\\&+3\left(-4+4g-2 E_0^2\right) \varphi_1^2  \varphi_4
+3 (-4+4 g) \varphi_3 \varphi_4^2
+3 (2-2 g) \varphi_3^2 \varphi_4
\\&+3\left(6-6g+E_0^2\right) \varphi_2 \varphi_3^2 
+3  \left(-8+8g-E_0^2\right)\varphi_2^2 \varphi_3
+3 \left(-2+2g-E_0^2\right)\varphi_2^2 \varphi_4 
\\&+6\left(2-2g+E_0^2\right)\varphi_1 \varphi_2 \varphi_4  
+6 \left(2-2g+E_0^2\right)\varphi_2 \varphi_3 \varphi_4 
\\
\hline \end{array}\right.
\end{align*}
\caption{The triple intersection form of the resolved I$_N^s$ model $\mathscr{T}$. The coefficient of $\varphi_i \varphi_j \varphi_k$ is 6 times the triple intersection number $D_i\cdot D_j\cdot D_k$ if $i,j,k$ are distinct, 3 times $D_i\cdot D_j\cdot D_k$ if only two of $i,j,k$ are the same, and equal to $D_i\cdot D_j\cdot D_k$ if $i=j=k$. $E_0$ is the genus $g$ curve in the base $B$ that supports the singular fiber. The cubic part of the prepotential $6\mathcal{F}(\varphi)$ matches with the triple intersection form $(\sum_i \varphi_i D_i)^3$  given the identification \eqref{mattercontent}. 
}\label{table:intersection}
\end{table}

\begin{table}[h!]
\begin{align*}
\left.\begin{array}{|c|c|}
\hline \text{Gauge Group} &\text{Prepotential}~~6\mathcal{F}(\varphi)  \\
\hline \,su(2)\, &  (8-8n_{\text{adj}} -n_F)\varphi_1^3\\
\hline \,su(3)\, &  (8-8n_{\text{adj}})\varphi_1^3+(8-8n_{\text{adj}}-n_F)\varphi_2^3
+3\left(-1+n_{\text{adj}}+c_{cl} -{n_F\over2}  \right)\varphi_1^2\varphi_2\\
&+3\left(-1+n_{\text{adj}}-c_{cl} +{n_F\over 2} \right)\varphi_1\varphi_2^2 \\
\hline\, su(4)\, &
 (8-8n_{\text{adj}}) \varphi_1^3
  + (8-8n_{\text{adj}}-n_F)\varphi_2^3 
  +(8-8n_{\text{adj}}-2n_A)\varphi_3^3  \\
& +3(c_{cl} -{n_F\over 2}+n_A ) \varphi_1^2 \varphi_2
+3(-2+2n_{\text{adj}} -c_{cl} +{n_F\over2} -n_A)\varphi_1\varphi_2^2\\
&
+3( -2n_A) \varphi_1^2\varphi_3+3(-2+2n_{\text{adj}}+c_{cl} +{n_F\over2} -n_A ) \varphi_2^2 \varphi_3\\
&
+3(-c_{cl} -{n_F\over2} +n_A)\varphi_2\varphi_3^2 
+ 6n_A\varphi_1\varphi_2\varphi_3\\
\hline \,su(5)\, & 
(8-8 n_{\text{adj}}) \varphi_1^3
+\left(8 - 8 n_{\text{adj}}- n_A\right)\varphi_2^3 
+ \left(8-8n_{\text{adj}}-n_F\right)\varphi_3^3\\
&
+ \left(8-8n_{\text{adj}}-2n_A\right)\varphi_4^3
+3  \left(-3+3n_{\text{adj}} -c_{cl} +{n_F\over 2} - {n_A\over2}\right) \varphi_1\varphi_2^2
\\&+3\left(1-n_{\text{adj}} +c_{cl} - {n_F\over 2} +{n_A\over2}\right)  \varphi_1^2 \varphi_2 
+3(- n_A)\varphi_2 \varphi_4^2 
+3\left(-2n_A\right) \varphi_1^2  \varphi_4
\\&+3 ( 1-n_{\text{adj}} -c_{cl} -{n_F\over2} +{3n_A\over2} ) \varphi_3 \varphi_4^2
+3 (-3+3n_{\text{adj}} +c_{cl} +{n_F\over2} -{3n_A\over2}) \varphi_3^2 \varphi_4
\\&+3\left(-1+n_{\text{adj}} - c_{cl} +{n_F\over2} -{n_A\over2}\right) \varphi_2 \varphi_3^2 
+3  \left(-1+n_{\text{adj}} +c_{cl} - {n_F\over2} +{n_A\over2}\right)\varphi_2^2 \varphi_3
\\&+3 \left(-n_A\right)\varphi_2^2 \varphi_4 
+6n_A\varphi_1 \varphi_2 \varphi_4  
+6 n_A\varphi_2 \varphi_3 \varphi_4 
\\
\hline \end{array}\right.
\end{align*}
\caption{The prepotential for the 5$d$ $su(N)$ gauge theory with $n_F$, $n_A$, $n_{\text{adj}}$ hypermultiplets in the fundamental, two-index antisymmetric, and adjoint representation, respectively. The prepotential $6\mathcal{F}(\varphi)$ matches with the triple intersection form $(\sum_i \varphi_i D_i)^3$ in Table \ref{table:intersection} given the identification \eqref{mattercontent}. We have ignored the quadratic terms in the prepotentials.
}\label{table:prepotential}
\end{table}

\subsection{The Euler Characteristic}\label{sec:euler}

Similarly we can compute the Euler characteristic of the resolved I$_N^s$ model $\mathscr{T}$ by pushing forward the intersection down to the base. Let $c_3(\mathscr{T})$ be the third Chern class of $\mathscr{T}$, then we can express the Euler characteristic as
\begin{align}
\chi = \pi_* \circ f_{1*} \circ\cdots \circ f_{N-1*} \left(
 c_3(\mathscr{T}) \cap [\mathscr{T}]  
\right).
\end{align}
We leave the explicit calculation to \S\ref{app:Euler} and present the results in Table \ref{table:euler} for small $N$. The Euler characteristic can be computed straightforwardly for any given $N$, and from direct inspections of the answers, we obtain a closed form formula:
\begin{align}\label{euler}
\chi&=\begin{cases}
-60K^2 +60(1-g) +24E_0^2,~~~~~~~~~~~~~~~~~~~~~~\text{if}~~~N=2,\\
-60K^2  +32N(1-g) +N(15-N)E_0^2,~~~~~~~~\text{if}~~~N\ge3.
\end{cases}
\end{align} 
This agrees with a  formula  proved in \cite{grassi2000group,Grassi:2011hq}. For similar techniques on computing the Euler characteristics, see \cite{Sethi:1996es,Klemm:1996ts,Aluffi:2007sx,Aluffi:2009tm,Esole:2011cn,fullwood2012stringy,Esole:2014dea}.

\begin{table}[h!]
\begin{align*}
\left.\begin{array}{|c|c|}
\hline  \text{Singular Fiber}&\text{Euler Characteristic}~ ~\chi \\
\hline \,\text{I}_2^s\, &  -60 K^2+60(1-g) +24E_0^2 \\
\hline \,\text{I}_3^s\, &  -60K^2 + 96(1-g) +36E_0^2 \\
\hline\, \text{I}_4^s\, & -60K^2 + 128(1-g) +44E_0^2\\
\hline \,\text{I}_5^s\, & -60K^2 + 160(1-g) +50E_0^2
\\
\hline \end{array}\right.
\end{align*}
\caption{The Euler characteristic of the resolved Calabi-Yau Weierstrass model with singular fiber of the type I$_N^s$. $E_0$ is the genus $g$ curve that supports the singular fiber in the base surface $B$. $K$ is the canonical class of the base $B$. 
}\label{table:euler}
\end{table}

\section{M-theory on I$_N^s$ Elliptic Calabi-Yau Threefolds}\label{sec:Mtheory}

In \S\ref{sec:kahler}, we discuss the relation between  the Coulomb branches of the 5$d$ theories obtained from M-theory compactification and the relative K\"ahler cones of the internal Calabi-Yau threefolds \cite{Witten:1996qb,Morrison:1996xf,Intriligator:1997pq,Aspinwall:2000kf} (see also \cite{Hayashi:2014kca}). For concreteness, we focus on the case with $su(N)$ gauge group, but the discussion is completely general for any gauge group. 
In \S\ref{sec:chmatter}, we match the triple intersection numbers of the Calabi-Yau threefolds with the 5$d$  gauge theory Chern-Simons levels (or equivalently, the prepotentials), and thereby determine the charged matter contents of these 5$d$ theories obtained from M-theory compactified on I$_N^s$ Weierstrass model with arbitrary algebraic base $B$.

\subsection{K\"ahler Cones and Coulomb Branches}\label{sec:kahler}

 The K\"ahler class of a  Calabi-Yau threefold $\mathscr{T}$ with all the singularities resolved depends on $b_2=\dim H^2(\mathscr{T})$ parameters.\footnote{To avoid extra supersymmetries, we assume $h^{2,0}(\mathscr{T})=0$.} Among these $b_2$ parameters, one combination is associated to a hypermultiplet which controls the size of $\mathscr{T}$. The remaining $b_2-1$ of them are associated to vector multiplets parametrizing the Coulomb branch of the 5$d$ theory \cite{Cadavid:1995bk}.

Let us  assume that a collection of surfaces $D_i\in H_4(\mathscr{T},\mathbb{Z})$ in $\mathscr{T}$ shrink to a curve $E_0$ when we approach a  singular point on the K\"ahler moduli space. The fiber of $D_i$ along the curve $E_0$ will be denoted by $\varepsilon_i\in H_2(\mathscr{T},\mathbb{Z})$, which shrinks to zero size in the above limit. To obtain an enhanced $su(N)$ gauge symmetry at the singular point on the K\"ahler moduli space, we assume that, over a generic point on $E_0$, the fibers $\varepsilon_i$ are $\mathbb{P}^1$'s intersecting with each other as in an affine ${su}(N)$ Dynkin diagram.

The resolved Weierstrass model I$_N^s$ reviewed in \S\ref{sec:Weierstrass} is a simple example of  general Calabi-Yau threefolds with enhanced $su(N)$ gauge symmetry described above. The curve $E_0$ is the curve in the base $B$ that supports the I$_N^s$ singular fiber.  $D_i$ is the   surface swept out by the $i$-th $\mathbb{P}^1$ in the I$_N^s$ fiber along the curve $E_0$ in the base $B$.

In M-theory compactification, the reduction of the three-form on the harmonic (1,1)-forms dual to $D_i$  give rise to 5$d$ Cartan gauge fields of ${su}(N)$. It is therefore natural to identify the coroot lattice of ${su}(N)$ as a sublattice of $H^2(\mathscr{T},\mathbb{Z})$. The dual lattice $H_2(\mathscr{T},\mathbb{Z})$ then contains the weight lattice of ${su}(N)$.  In our convention, the intersection product $D_i\cdot \varepsilon_j$ is the \textit{negative} of the evaluation of coroots on weights:
 \begin{align}
 D_i ( \varepsilon_j) = -D_i \cdot \varepsilon_j.
 \end{align}

Among the classes in $H_2(\mathscr{T},\mathbb{Z})$, we have the fiber classes $\varepsilon_i$ of $D_i$, which  correspond to the simple roots of ${su}(N)$. M2-branes wrapping around $\varepsilon_i$ then give rise to massive vector multiplets ($W$-bosons) or adjoint hypermultiplets whose masses are proportional to the sizes of the cycles \cite{Witten:1996qb}. The charges of these states are given by minus of  the  intersection numbers between $D_i$ and $\varepsilon_j$.  
In addition to $\varepsilon_i$, we also have the classes $\sigma_k$ ($k=1,\cdots,N)$ in $H_2(\mathscr{T},\mathbb{Z})$ corresponding to the weights in the fundamental representation of ${su}(N)$ for each fundamental hypermultiplet. From the identification with the weights in the fundamental representation, we have $\sum_{i=1}^N \sigma_i=0$. They are related to $\varepsilon_i$ by
\begin{align}
\varepsilon_i= \sigma_i - \sigma_{i+1}.
\end{align}

Now fix a K\"ahler class $\varphi\in H^2(\mathscr{T})$. This in turn fixes  the vev of the real scalars in the vector multiplets of the low energy 5$d$ theory. Since the simple roots $\varepsilon_i$ are represented by effective curves, we have
\begin{align}
\int_{\varepsilon_i}\varphi>0.
\end{align}
From $\varepsilon_i =\sigma_i -\sigma_{i+1}$, this implies
\begin{align}
\int_{\sigma_i}\varphi >\int_{\sigma_{i+1}} \varphi.
\end{align}
Since $\sum_{i=1}^N\sigma_i=0$, we have the following constraint on the areas of the curves $\sigma_i$:
\begin{align}
\sum_{i=1}^N\int_{\sigma_i}\varphi=0.
\end{align}
Combining the above two facts, we learn that there is an integer $\ell$ between 1 and $N-1$ such that
\begin{align}
\begin{split}
&\int_{\sigma_i} \varphi>0,~~~i=1,\cdots,\ell,\\
&\int_{\sigma_j} \varphi<0,~~~j=\ell+1,\cdots,N.
\end{split}
\end{align}
That is, there must exist an $\ell$ such that 
 $\sigma_1,\cdots,\sigma_\ell$ and $-\sigma_{\ell+1},\cdots, -\sigma_N$ are all effective curves.

Now as we vary $\varphi$ towards the boundary of the relative K\"ahler cone,  there exists some effective curves whose areas approach zero. Let us consider the boundary of the relative K\"ahler cone where 
\begin{align}\label{boundary}
\int_{\sigma_\ell} \varphi=0.
\end{align}
As we cross this boundary, we enter into the relative K\"ahler cone of another Calabi-Yau which is related to the original one by a flop. 

Let us make the connection to the gauge theory side. Notice that the boundary defined by \eqref{boundary} is precisely the walls \eqref{wall} on the Coulomb branch where some of the hypermultiplets become massless. Thus we have the following correspondence:
\begin{align*}
\text{boundary of the relative K\"ahler cone:}~\int_{\sigma_\ell} \varphi=0 \,\leftrightarrow \, \text{wall on the Coulomb branch:}~\varphi(w_\ell) = 0
\end{align*}
where $w_\ell$ denotes the weight corresponding to $\sigma_\ell$. This correspondence naturally comes from the identification between the intersection numbers and the evaluation of coroots on weights.

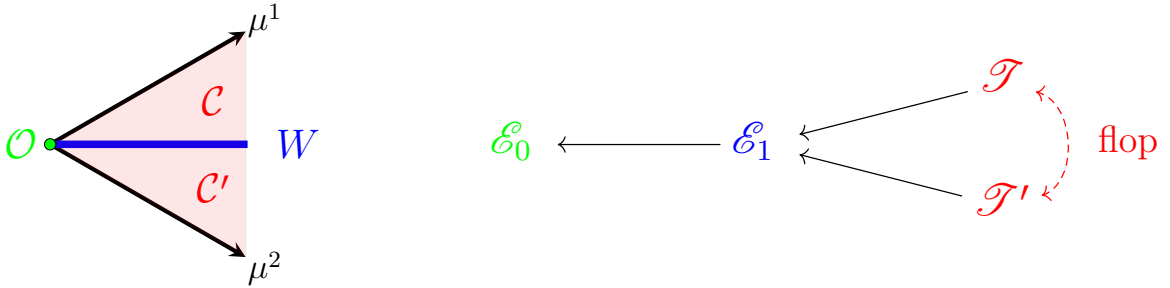
\begin{figure}[h!t]

\begin{tabular}{lcr}
\raisebox{-2.cm}{
\begin{tikzpicture}[scale=1.5]
\coordinate (O) at (0,0);
\coordinate (A1)  at (30:2);
\coordinate (A2)  at (-30:2);
\draw[ultra thick, -stealth,black] (O)--(A1);
\draw[ultra thick, -stealth,black] (O)--(A2);
\draw[line width=1mm,blue] (O)--(0:1.75);
\draw[fill=red, opacity=.1] (O)--(A1)--(A2);
\draw[fill=green, opacity=1] (0,0) circle (.05);
\node at (-.2,0) {\color{green!} \scalebox{1.3}{$\mathcal{O}$~}};
\node at (15:1.5) {\color{red} \scalebox{1.3}{$\mathcal{C}$}};
\node at (-15:1.5) {\color{red} \scalebox{1.3}{$\mathcal{C}'$}};
\node at (30:2.2) {\color{black} \scalebox{1.1}{$\mu^1$}  };
\node at (-30:2.2) {\color{black} \scalebox{1.1}{$\mu^2$}};
\node at (0:1.9) {\color{blue} \scalebox{1.3}{~~~~~$W$}};
\end{tikzpicture}}

\phantom{~~~~~~~~~~}

\begin{tikzcd}[column sep=huge, row sep=tiny,scale=1.5]
& & \scalebox{1.4}{\color{red}$ \mathscr{T}$}  \arrow[bend left=75, leftrightarrow, dashed, color=red]{dd}[right] {\quad \text{\large flop}}\\{\color{green}
 \scalebox{1.4}{ $\mathscr{E}_0$ }}\arrow[leftarrow]{r} & \scalebox{1.4}{{\color{blue}$ \mathscr{E}_1$} }\arrow[leftarrow]{ur} \arrow[leftarrow]{dr}& \\\
&    &  \scalebox{1.4}{{\color{red}$\mathscr{T}'$}}
\end{tikzcd} 

\end{tabular}

\caption{Left: The $su(3)$ Coulomb branch. The Coulomb branch is divided by the line $W_{w_2}$ into two subchambers $\mathcal{C}$ and $\mathcal{C}'$. The line $W$ is the codimension one \textit{wall} where the Coulomb-Higgs branch intersects the Coulomb branch. Right: The network of small resolutions for the I$_3^s$ model. Each letter stands for a (partial) resolution of the original singular Weierstrass model $\mathscr{E}_0$ and each arrow represents a blowup. By going along (against) an arrow, we blow down (up) a variety.
 The identifications between the Coulomb branch with the (partially) resolved varieties are given by $\color{red}\mathscr{T} = \mathcal{C}$, $\color{red}\mathscr{T}' = \mathcal{C}'$, $\color{blue}\mathscr{E}_1= W$, and $\color{green}\mathscr{E}_0=\mathcal{O}$. The flop is realized as the reflection with respect to the line (wall) $W$.}\label{intro3}

\end{figure}

Let us summarize the discussion so far. In M-theory compactification, the K\"abler class (except for one modulus that controls the size of the Calabi-Yau) is identified as the vev of the real scalars in the vector multiplet. Therefore the subchambers on the Coulomb branch naturally corresponds to the relative K\"abler cones of the Calabi-Yau. The walls on Coulomb branch where some hypermultiplet scalars become massless correspond to the boundary of the relative K\"abler cone where some effective curves shrink. The cases for $su(3)$ gauge theory is illustrated in Figure \ref{intro3}.

\subsection{5$d$ Charged Matter Contents}\label{sec:chmatter}

We have reviewed how the \textit{topology} of the Coulomb branch (\textit{e.g.} division into different subchambers) of the 5$d$ gauge theories can be seen from the relative K\"ahler cones of the resolved  Calabi-Yau. The main goal of the present paper is to reproduce the \textit{metric} on the Coulomb branch, \textit{i.e.} the prepotential of the Coulomb branch effective action, from the triple intersection  numbers $D_i\cdot D_j\cdot D_k$ in the resolved elliptic Calabi-Yau obtained in \cite{Esole:2014bka}. Our calculation serves as an explicit demonstration of the general scenario of M-theory compactified on Calabi-Yau threefolds \cite{Witten:1996qb,Morrison:1996xf,Intriligator:1997pq}.

Let $\varphi\in H^2(\mathscr{T})$ be a K\"ahler class of the Calabi-Yau threefold $\mathscr{T}$. In the low energy limit of M-theory compactified on $\mathscr{T}$, the K\"ahler moduli are controlled by the Cartan scalars in the vector multiplets (except for the modulus controlling the size of $\mathscr{T}$, which belongs to a hypermultiplet), and we will use the same symbol $\varphi$ to denote these scalars. The triple intersection numbers are identified as the 5$d$ Chern-Simons levels $c_{ijk}$  \eqref{CS},
\begin{align}
c_{ijk} = D_i \cdot D_j \cdot D_k.
\end{align}
Or equivalently, the triple intersection form is identified as the prepotential of the 5$d$ gauge theory,
\begin{align}\label{identification}
6\mathcal{F} (\varphi)= 
(\sum_{i=1}^{N-1} \varphi_i D_i)^3
\end{align}
where $D_i\in H_4(\mathscr{T},\mathbb{Z})$ are the surfaces swept out by the $i$-th $\mathbb{P}^1$ in the singular fiber along the curve $E_0$ in the base. $D_i$ corresponds to the simple coroots of ${su}(N)$. The product on the right-hand side is given by the intersection product. 

From the above identification we can determine the charged matter contents $n_\mathbf{R}$ and the bare Chern-Simons level $c_{cl}$ of the 5$d$  theory obtained by compactifying M-theory on a resolved elliptic Calabi-Yau  with singular fiber I$_N^s$. To begin with,  the number of adjoint hypermultiplets $n_{\text{adj}}$ is simply given by the genus $g$ of the curve $E_0$. 
\begin{align}
n_{\text{adj}}=g.
\end{align}
This can be seen from explicitly quantizing the moduli space of the 4-supercharge quantum mechanics from a wrapped M2-brane \cite{Witten:1996qb} (see also \cite{Katz:1996ht}).

For the other quantities, we will discuss the I$_2^2$ and I$_3^s$ cases separately from the general $N$ case, since the antisymmetric representation is special (or absent) for $su(2)$ and $su(3)$.

\subsubsection{The I$_2^s$ Model}
In the I$_2^s$ model, the triple intersection of $D_1$ is (see Table \ref{table:intersection})
\begin{align}\label{su2triple}
D_1^3 =  -8 +8g -6E_0^2,
\end{align}
where $E_0^2$ is the self-intersection of the curve $E_0$ in the base $B$.  The triple intersection form is
\begin{align}\label{int2}
(\text{I}_2^s):~~(\sum_i \varphi_iD_i )^3  =(  -8 +8g -6E_0^2)\varphi_1^3.
\end{align}
Matching this with the 5$d$ gauge theory prepotential \eqref{E1} $6\mathcal{F}(\varphi) = (8 - 8n_{\text{adj}} -n_F)\varphi_1^3$, we obtain the number of  hypermultiplets in the fundamental representation,
\begin{align}\label{su2matter}
n_F = 16-16g+6E_0^2,~~~~~n_{\text{adj}}=g.
\end{align}
Note that there is no antisymmetric representation for $su(2)$. There is no classical Chern-Simons level for the $su(2)$ theory because the third-order Casimir is trivial.

\subsubsection{The I$_3^s$ Model and Flop Transitions}

There are two resolutions $\mathscr{T},\, \mathscr{T}'$ for the I$_3^s$ model (see Figure \ref{tree3}). The triple intersection form of the resolution $\mathscr{T}$ is (see Table \ref{table:intersection})
\begin{align}\label{int3}
\begin{split}
(\text{I}_3^s)~~~\mathscr{T}:~~(\sum_i\varphi_iD_i)^3=& (8-8g)\varphi_1^3+(-10+10g -6E_0^2)\varphi_2^3+3( -10+10g -3E_0^2)\varphi_1^2\varphi_2\\
&+3( 8-8g +3E_0^2)
\varphi_1\varphi_2^2.
\end{split}
\end{align}
For the resolution $\mathscr{T}'$, the classes of the centers of the blowups (shown in Figure \ref{tree3}) are the same since $[s]= [y+a_1x+a_{3,1}e_0]=[y]$. The only difference lies in the classes for the surfaces $D_i$ in \eqref{Dclass}. It follows  that the triple intersection numbers  in $\mathscr{T}'$ are obtained by exchanging $D_1$ with $D_2$ from those in $\mathscr{T}$,
\begin{align}
\begin{split}
(\text{I}_3^s)~~~\mathscr{T}':~~(\sum_i\varphi_iD_i)^3=&(-10+10g -6E_0^2)\varphi_1^3+ (8-8g)\varphi_2^3+3( 8-8g +3E_0^2)
\varphi_1^2\varphi_2\\
&+3( -10+10g -3E_0^2)\varphi_1\varphi_2^2.
\end{split}
\end{align}

The triple intersection numbers above in the resolutions $\mathscr{T}$ and $\mathscr{T}'$ exactly matches with the gauge theory prepotentials \eqref{l=2} and \eqref{l=1} in the two subchambers. The jumps in the triple intersections under the flop transition between $\mathscr{T}$ and $\mathscr{T}'$ are precisely captured by the discontinuities of the gauge theory prepotentials (or equivalently, the 5$d$ Chern-Simons levels).

Matching the triple intersection numbers with the prepotentials in either subchamber, we determine the number of hypermultiplets in the fundamental representation $n_F$ and the classical Chern-Simons level $c_{cl}$ to be
\begin{align}\label{su3matter}
n_F= 18-18g+6E_0^2,~~~~~n_{\text{adj}}=g,~~~~~~~c_{cl}=0.
\end{align}
Note that the antisymmetric representation of $su(3)$ is the same as antifundamental representation. Since  representations come in conjugate pairs in a hypermultiplet, there is no distinction between hypermultiplets in the  fundamental and antifundamental representations, and we could as well call both of them as in the fundamental representation.

\subsubsection{The I$_N^s$ Model}

 By comparing the triple intersection form in Table \ref{table:intersection} with the gauge theory prepotential in Table \ref{table:prepotential}, we obtain the charged matter contents for the 5$d$ $su(N)$ gauge theory from M-theory compactified on the resolved I$_N^s$ model $\mathscr{T}$:
\begin{align}\label{mattercontent}
\begin{split}
&n_F = 16-16g +(8-N) E_0^2, \\
&n_A= 2-2g+E_0^2,\\
&n_{\text{adj}}=g,\\
&c_{cl} = 0,
\end{split}
\end{align}
where $E_0$ is the genus $g$ curve in the base $B$ that supports the singular fiber and $E_0^2$ is its the self-intersection in $B$.  It should be emphasized that our result \eqref{mattercontent} holds for any base $B$  (compact or not) and  any non-singular curve $E_0$ in $B$.

In fact, the numbers of hypermultiplets $n_F,n_A,n_{\text{adj}}$ \eqref{mattercontent} from M-theory on the elliptic Calabi-Yau $\mathscr{T}$ are identical to those of the 6$d$ theories arising from F-theory compactified on the same Calabi-Yau\cite{Sadov:1996zm}.
 Furthermore, they are the only anomaly-free 6$d$ theories  with a single $su(N)$ gauge group and  fundamental, antisymmetric, and adjoint hypermultiplets. 
 The 6$d$ origins of these 5$d$ theories also explain why the bare  Chern-Simons level is zero $c_{cl}=0$.\footnote{For 5$d$ theories coming from circle reduction of  6$d$ $\mathcal{N}=(1,0)$ theories, the gauge field has three different kinds of 6$d$ origins. One is the Kaluza-Klein gauge field, denoted by $A_G$, while the others either come from the 6$d$ tensor fields or the 6$d$ gauge fields, denoted collectively by $A_T$ and $A_V$, respectively. The Green-Schwarz-West-Sagnotti mechanism in 6$d$ induces a 5$d$ bare mixed Chern-Simons term $A_T\wedge F_V\wedge F_V$ (see, for example, \cite{Bonetti:2013cza,Ohmori:2014kda}). However, there is no bare   Chern-Simons term of the form $A_V\wedge F_V\wedge F_V$ induced from 6$d$, hence $c_{cl}=0$.  } We will review the anomaly cancellation equations in \S \ref{sec:anomaly}.

\section{6$d$ Anomaly Cancellation}\label{sec:anomaly}

In \S\ref{sec:cancel}, we review the gauge and gravitational anomalies in 6$d$ and the Green-Schwarz-West-Sagnotti mechanism for anomaly cancellation.  In \S\ref{sec:gaugeanomaly}, we show that  the 5$d$ charged matter contents \eqref{mattercontent} are lifted to the most general 6$d$ theories that can solve the gauge and mixed anomaly cancellation equations, with a single $su(N)$ gauge group and fundamental, antisymmetric, and adjoint representations. In \S\ref{sec:gravanomaly}, we consider 6$d$ theories that are coupled to gravity and check the cancellation of the pure gravitational anomalies from our explicit answers for the Euler characteristics obtained in \S\ref{sec:euler}.

\subsection{Green-Schwarz-West-Sagnotti Mechanism}\label{sec:cancel}

We first review the anomalies in 6$d$ $\mathcal{N}=(1,0)$ theory. 
Suppose a 6$d$ theory has an anomalous symmetry transformation $\delta_\Lambda$, under which the action $S$ is not invariant, \textit{i.e.} $\delta_\Lambda S = \int I_6(\Lambda)$.
We can conveniently encode this anomaly $I_6(\Lambda)$ by a 8-form  polynomial $I_8(R,F)$ defined by the descent equations:
\begin{align}\label{I8}
\delta_\Lambda I_7 = dI_6(\Lambda),~~~I_8 = dI_7.
\end{align}
The anomaly 8-form for a 6$d$ $\mathcal{N}=(1,0)$  theory is given in terms of the characteristic classes for the gauge fields and the gravitational fields \cite{AlvarezGaume:1983ig,Erler:1993zy}:\footnote{We assume that there are no $u(1)$ factors and no hypermultiplet charged under more than two simple groups.} 
\begin{align}
\begin{split}
I_8& ={1\over 360} (H-V-273+29T) \left[
\tr R^4  +{5\over 4} (\tr R^2)^2\right]\\
& -{2\over 3} \sum_I X_I^{(4)}
+ {9-T \over 8 } (\text{tr} R^2 )^2 + {1\over 6} \text{tr}R^2 \sum_I  X_I^{(2)}+4\sum_{I<J} Y_{IJ},
\end{split}
\end{align}
where 
\begin{align}
&X_I^{(n)} = \tr_{\text{adj}}  F_I^n - \sum_{\mathbf{R}} \,  n^I_{\mathbf{R}} \tr_{\mathbf{R}} F^n_I,\\
&Y_{IJ}=\sum_{\mathbf{R},\mathbf{R}'}\,  n^{IJ}_{\mathbf{R} ,\mathbf{R}'} \,  \tr_{\mathbf{R}}F^2_I\, \tr_{\mathbf{R}'} F^2_J.
\end{align}
We denote by $\tr$ the trace in a preferred representation, which for $su(N)$ is the fundamental representation. The sum in $I,J$ runs over all the gauge groups. $H, V, T$ are the numbers of hypermultiplets, vector multiplets, and tensor multiplets, respectively. $n_{\mathbf{R}}^I$ is the number of hypermultiplets in the representation $\mathbf{R}$ under the $i$-th gauge group. $n_{\mathbf{R}, \mathbf{R}'}^{IJ} $ is the number of hypermultiplets in the bifundamental representation $(\mathbf{R}, \mathbf{R}')$ under the $I$- and $J$-th gauge group.

The  anomaly can be canceled by the Green-Schwarz-West-Sagnotti mechanism \cite{Green:1984bx,Sagnotti:1992qw} if the  8-form can be written as
\begin{align}\label{GSS}
I_8  ={1\over 2} \Omega_{\alpha\beta} X^\alpha X^\beta,
\end{align}
where $\Omega_{\alpha\beta}$ is a symmetric bilinear form on the vector space $\mathbb{R}^{1,T}$ for the tensor fields, including the one in the gravity multiplet.
 Here $X^\alpha$ is a 4-form that can be written as
\begin{align}
X^\alpha = {1\over 2} a^\alpha \tr R^2 + \sum_I {2\over\lambda_I} b_I^\alpha \tr \, F_I^2,
\end{align}
for some coefficients $a^\alpha$ and $b_I^\alpha$, which are vectors on $\mathbb{R}^{1,T}$. $\lambda_I$ is a normalization factor given below:
\begin{align*}
\left.\begin{array}{|c|c|c|c|c|c|c|c|c|c|}\hline \mathfrak{g} & A_n & B_n & C_n & D_n & E_6 & E_7 & E_8 & F_4 & G_2 \\\hline \lambda & 1 & 2 & 1 & 2 & 6 & 12 & 60 & 6 & 2 \\\hline \end{array}\right.
\end{align*}
 We will often use $\cdot$ as the inner product on  $\mathbb{R}^{1,T}$. For example,
\begin{align}
a \cdot b_I = \Omega_{\alpha\beta} \,a^\alpha b_I^\beta.
\end{align}

The general anomaly 8-form \eqref{I8} can be rearranged into the form \eqref{GSS} if the matter contents and the gauge groups satisfy 
\begin{align}\label{gravanomaly}
\begin{split}
&H-V=273-29T,\\
&a\cdot a  =  9-T,
\end{split}
\end{align}
and
\begin{align}\label{gaugeanomaly}
\begin{split}
&B_{\text{adj}} = \sum_{\mathbf{R} } \, n_{\mathbf{R}}^I B^I_{\mathbf{R}},\\
& a\cdot b_I = {\lambda_I\over 6}\left( A^I_{\text{adj}} -\sum_{\mathbf{R}} n^I_{\mathbf{R}}A^I_{\mathbf{R}}\right),\\
&b_I  \cdot b_I =- {\lambda_I^2\over 3}\left( C^I_{\text{adj}} -\sum_{\mathbf{R}} n^I_{\mathbf{R}}C_{\mathbf{R}}^I\right),\\
&b_I\cdot b_J  = \lambda_I\lambda_J\sum_{\mathbf{R},\mathbf{R}'}n^{IJ}_{\mathbf{R},\mathbf{R}'} \, A^I _\mathbf{R} A^J _{\mathbf{R}'},
\end{split}
\end{align}
for some choices of $a^\alpha$ and $b_I^\alpha$. There is no sum in $I$ and $J$. 
These equations are known as the anomaly cancellation equations. 
Here the coefficients $A_\mathbf{R}, B_\mathbf{R},C_\mathbf{R}$ are defined by the following relations between traces in different representations:
\begin{align}\label{ABC}
&\tr_\mathbf{R}F^2 =A_\mathbf{R}\tr F^2,\\
&\tr_\mathbf{R}F^4 =B_\mathbf{R}\tr F^4 + C_\mathbf{R} (\tr F^2)^2.
\end{align}
For $su(N)$, the coefficients are 
\begin{align*}
\begin{split}
\left.\begin{array}{|c|c|c|c|}\hline \mathbf{R} & A_\mathbf{R} & B_\mathbf{R} & C_\mathbf{R} \\\hline \text{Fundamental} & 1 & 1 & 0 \\\hline \text{Antisymmetric} & N-2 & N-8 & 3 \\\hline \text{Adjoint} & 2N & 2N & 6 \\\hline \end{array}\right.
\end{split}
\end{align*}

\subsection{Gauge and Mixed Anomalies}\label{sec:gaugeanomaly}

For the theory with a single $su(N)$ vector multiplet, $n_F$ fundamental hypermultiplets, $n_A$ antisymmetric hypermultiplets, $n_{\text{adj}}$ adjoint hypermultiplets (and possibly a gravity multiplet and  some numbers of tensor multiplets), the gauge and mixed anomaly cancellation equations \eqref{gaugeanomaly} determine the charged matter contents to be \cite{Sadov:1996zm,Kumar:2009ac}
\begin{align}\label{6dmatter}
\begin{split}
&n_F =- 8a\cdot b - N b\cdot b,\\
&n_A = -a\cdot b,\\
&n_{\text{adj}} = 1 + {1\over 2} \left(a\cdot  b + b\cdot b \right).
\end{split}
\end{align}
Comparing the 6$d$ matter contents \eqref{6dmatter} obtained from anomaly cancellation with the 5$d$ matter contents \eqref{mattercontent} obtained from M-theory compactified on elliptic Calabi-Yau threefolds, we find an exact agreement if we identify
\begin{align}\label{6d5d}
g= 1 + {1\over 2} (a\cdot b + b\cdot b),~~~~E_0^2 = b\cdot b,
\end{align}
where recall that $E_0$ is the curve in the base of the elliptic Calabi-Yau threefold that supports the type I$_N^s$ singular fiber, and $g$ is the genus of the curve $E_0$. Indeed, for 6$d$ theories from F-theory compactification, the coefficients $a^\alpha$ and $b^\alpha$ are identified as the canonical class $K$ of the base and the curve $E_0$,
\begin{align}
a^\alpha\rightarrow K,~~~~b^\alpha\rightarrow E_0,
\end{align}
 with the inner product $\cdot$  identified as the intersection product \cite{Sadov:1996zm} (see also \cite{Kumar:2010ru}). This justifies the relation \eqref{6d5d} where the first equation is simply the adjunction formula $-KE_0 -E_0^2 = 2-2g$.

To conclude, we have shown that  all the 5$d$ theories  \eqref{mattercontent}  from M-theory  on  I$_N^s$ elliptic Calabi-Yau threefolds, which we determined from triple intersection numbers,  can be lifted to  6$d$ $\mathcal{N}=(1,0)$ theories while satisfying the gauge and mixed anomaly cancellation equations. These 6$d$ parent theories have their natural origins from F-theory compactifications on the same elliptic Calabi-Yau threefolds, and therefore our results serve as a direct check of the F/M-theory duality (see Figure \ref{fig:flowchart}).

\subsection{Gravitational Anomalies and Euler Characteristics}\label{sec:gravanomaly}

In this subsection we will switch the gear to check the pure gravitational anomaly cancellation for 6$d$ theories obtained from F-theory compactified on the compact I$_N^s$ Weierstrass models with general bases $B$, following the spirit in \cite{grassi2000group,Grassi:2011hq}. The key for this calculation is the explicit results on the Euler characteristics of the I$_N^s$ Weierstrass models in \S\ref{app:Euler}. 

First of all, the second equation in \eqref{gravanomaly} fixes the number of tensor multiplets to be
\begin{align}
T=9-K^2,
\end{align}
where we have identified the vector $a$ as the canonical class $K$ in the lattice $H^2(B,\mathbb{Z})$.

Next, we will  determine the number of 6$d$ neutral hypermultiplets $H_0$.  This can be determined from a detour to 5$d$ by compactifying the theory on a circle and go to the Coulomb branch. The number of 5$d$ massless hypermultiplets $H_{m=0}$ on the Coulomb branch is \cite{Cadavid:1995bk,Vafa:1996xn}
\begin{align}
H_{m=0} = h^{2,1}(\mathscr{T}) +1,
\end{align}
where the 1 comes from the hypermultiplet controlling the overall size of $\mathscr{T}$. 

Let us relate the  neutral hypermultiplets $H_0$  to  the 5$d$  massless hypermultiplets $H_{m=0}$  on  the Coulomb branch. Obviously each neutral hypermultiplet remains massless when moving on to the Coulomb branch. However, not every massless hypermultiplet comes from a neutral hypermultiplet at the origin of the Coulomb branch. Indeed, a  hypermultiplet state with trivial weight  in a nontrivial representation can contribute to $H_{m=0}$ when moving on to the Coulomb branch. For example, there are $N-1$ trivial weights in the adjoint representation of $su(N)$. 

In the case of interest, we only have fundamental, antisymmetric, and adjoint representation, thus we have
\begin{align}\label{preH0}
H_0 = H_{m=0} - \text{rk}(G)\, n_{\text{adj}} = h^{2,1}(\mathscr{T}) +1- \text{rk}(G) \,g,
\end{align}
where we have used $n_{\text{adj}} =g$. For our case, $G=su(N)$.

On the other hand, the number of $u(1)$ vector multiplets $V^{5d}$ on the Coulomb branch is given by  \cite{Cadavid:1995bk}
\begin{align}\label{V5d}
V^{5d} =  h^{1,1}(\mathscr{T})-1,
\end{align}
where the -1 is because the overall size of $\mathscr{T}$ is in a hypermultiplet. The $u(1)$ vector multiplets in 5$d$ can be categorized in the following three groups by their 6$d$ origins:
\begin{itemize}
\item One of them  from the 6$d$ gravity multiplet.
\item $T$ of them from the 6$d$ tensor multiplets.
\item rk$(G)$ of them from the Cartan parts of the 6$d$ vector multiplets.
\end{itemize}
Thus we have $V^{5d} =1+T+ \text{rk}(G)$. Together with \eqref{V5d}, $h^{1,1}(\mathscr{T})$ is determined to be
\begin{align}\label{h11}
h^{1,1}(\mathscr{T}) = 2+T+\text{rk}(G).
\end{align}

Using $\chi = 2h^{1,1}(\mathscr{T}) -2h^{2,1}(\mathscr{T})$ and \eqref{h11}, we can now express the number of 6$d$ neutral hypermultiplets \eqref{preH0} as
\begin{align}
H_0 = 3+T-{\chi\over 2}  +\text{rk}(G) \,(1- g),
\end{align}
with the Euler characteristics  given in \eqref{euler}. 

Finally, the number of 6$d$ charged hypermultiplets and 6$d$ vector multiplets are
\begin{align}
&H_{\text{ch}} = N n_F+ {N\choose 2}n_A + (N^2-1)n_{\text{adj}},\label{Hch}\\
&V=  N^2-1,
\end{align}
with $n_F$, $n_A$, and $n_{\text{adj}}$ given in \eqref{mattercontent} (or \eqref{6dmatter}) for $N\ge4$ and in \eqref{su2matter} and \eqref{su3matter} for $N=2$ and $N=3$, respectively. With all the above preparation, one can straightforwardly check that the first pure gravitational anomaly cancellation equation \eqref{gravanomaly} is satisfied,
\begin{align}
H_0 + H_{\text{ch}} -V = 273-29T.
\end{align}

If we reverse the logic and \textit{assume} that the pure gravitational anomaly  cancellation equations \eqref{gravanomaly} are satisfied, we can derive a simple expression for the Euler characteristic for the resolved Weierstrass model with singular fiber of type I$_N^s$ as in  \cite{grassi2000group,Grassi:2011hq}
\begin{align}\label{grassi}
\begin{split}
\chi &= -60K^2 +2H_{\text{ch}} -2 \big[\dim G-(1-g)\text{rk}(G)\big]\\
&=\begin{cases}
-60K^2 +60(1-g) +24E_0^2,~~~~~~~~~~~~~~~~~~~~~~\text{if}~~~N=2,\\
-60K^2  +32N(1-g) +N(15-N)E_0^2,~~~~~~~~\text{if}~~~N\ge3,
\end{cases}
\end{split}
\end{align}
which agrees with our answer in \eqref{euler}.\footnote{Note that the notion of ``charged" hypermultiplet $H_{\text{ch}}$ can be ambiguous in different references. Some authors (for example, in \cite{grassi2000group,Grassi:2011hq}) prefer to call a hypermultiplet ``charged" if it transforms nontrivially  under the \textit{Cartan} parts of the gauge symmetry transformation, \textit{i.e.} carrying nontrivial weights. In this paper, we call a hypermultiplet ``charged" if it belongs to a nontrivial representation of the whole gauge group. In particular, we include the Cartan parts of the adjoint hypermultiplets to the category of ``charged" hypermultiplets. Due to this ambiguity, our $H_{\text{ch}}$ in \eqref{Hch} differs from that in \cite{grassi2000group,Grassi:2011hq} by $g\,$rk$(G)$, \textit{i.e.} $H^{\text{here}}_{\text{ch}} = H^{\text{there}}_{\text{ch}} + g\, \text{rk}(G)$.}

\section{Outlook on Triple Intersection}\label{sec:outlook}

In previous sections we explicitly computed the triple intersection numbers to check the  F/M-theory duality. In this section, we  turn the logic around to predict the triple intersection numbers in elliptic Calabi-Yau threefolds by assuming the F/M-theory duality. This is similar to the approach in \cite{Intriligator:1997pq} for general Calabi-Yau threefolds, where the triple intersection numbers are related to the charged matter contents $n_\mathbf{R}$ from M-theory compactification. However, in the case of \textit{elliptic} Calabi-Yau threefolds, 
we gain extra information from F-theory compactification via the F/M-theory duality.

The duality chain is illustrated in Figure \ref{fig:flowchart}. Given a Weierstrass model $\mathscr{E}_0$ with a simple gauge group $G$ supported on the curve $E_0$,  using anomaly cancellation equations, we first determine the multiplicities $n_\mathbf{R}$ of charged hypermultiplets  of the low energy 6$d$ theories obtained from F-theory compactifcation. Next, we  compactify  the 6$d$ theories on a circle, and move on to the 5$d$ Coulomb branches. Finally, the Coulomb branch effective action is  given in \eqref{prepotential}, whose cubic coefficients are then the triple intersection numbers. 

Explicitly, the algorithm is given as follows:
\begin{itemize}
\item \textbf{Step 1}  ~Determine the representations $\mathbf{R}$ (without multiplicities) for the gauge group  $G$.
\item \textbf{Step 2}  ~Solve the multiplicities $n_{\mathbf{R}}$ of the charged hypermultiplets in terms of the intersection data in the base $B$ from the  gauge and mixed anomaly cancellation equations:
\begin{align}\label{Fanomaly}
\begin{split}
&B_{\text{adj}} = \sum_{\mathbf{R} } \, n_{\mathbf{R}}B_{\mathbf{R}},\\
& K\cdot E_0 = {\lambda\over 6}\left( A_{\text{adj}} -\sum_{\mathbf{R}} n_{\mathbf{R}}A_{\mathbf{R}}\right),\\
&E_0  \cdot E_0=- {\lambda^2\over 3}\left( C_{\text{adj}} -\sum_{\mathbf{R}} n_{\mathbf{R}}C_{\mathbf{R}}\right),
\end{split}
\end{align}
where $K$ is the anticanonical class of the base. $\lambda, A_\mathbf{R},B_\mathbf{R},C_\mathbf{R}$ are defined in \S\ref{sec:anomaly}. 
\item \textbf{Step 3} ~Let $\varphi$ be in the Weyl chamber $\mathfrak{h}/W_G$. The triple intersection form is then given by 
\begin{align}
\sum_{i,j,k=1}^{\text{rk}(G)} D_i \cdot D_j \cdot D_k \, \varphi_i \varphi_j \varphi_k=
{1\over 2} \left(
\sum_{\alpha\in G} |\varphi(\alpha)|^3 -\sum_\mathbf{R} n_\mathbf{R} \sum_{w\in \mathbf{R}} | \varphi(w)|^3\right),
\end{align}
with $n_\mathbf{R}$ obtained in Step 2. The sum in $\alpha$ is over all the roots of $G$ and the sum in $w$ is over all the weights of $\mathbf{R}$. The sign choices above reflect the jumps in the triple intersection numbers for different resolutions of the singular Weierstrass model.
\end{itemize}
There are two important subtleties in the algorithm above.  First, a complete answer for the matter representations $\mathbf{R}$ is still lacking for general elliptic Calabi-Yau threefolds with gauge groups supported on singular curves \cite{Morrison:2011mb}. Second, even given the distinct representation types in Step 1, the gauge and mixed anomaly cancellation equations can not in general determine the multiplicities $n_\mathbf{R}$ completely. This is because there are only three anomaly cancellation equations in \eqref{Fanomaly}, while one could in principle have more distinct representations $\mathbf{R}$.\footnote{One simple example is to take $G=SU(N)$ and the curve $E_0$ to have double point singularities. In this case the possible representations are the fundamental, antisymmetric, adjoint, and symmetric representations, and the anomaly cancellation equations alone cannot determine the multiplicities completely \cite{Sadov:1996zm}.}

In this paper, we checked the above proposal by explicit geometric calculations in the I$_N^s$ Weierstrass models (\textit{i.e.} $G=SU(N)$)   with nonsingular curves $E_0$ in arbitrary algebraic surfaces $B$ (compact or not).  The possible representations are assumed to be only the fundamental, antisymmetric, and adjoint representations, which are the most generic representations from codimension two singularities \cite{Katz:1996xe} and from quantizing the wrapped M2-brane quantum mechanics \cite{Witten:1996qb}. There is a straightforward generalization  to the cases with multiple gauge group factors, and it would be interesting to study the validity and the generality of this algorithm.

\subsection*{Example: The I$_2^s$ Model}

Let us illustrate the above algorithm in the simplest example, the I$_2^s$ Weierstrass model. In Step 1, the representations are assumed to be the fundamental and adjoint representations. In Step 2,  the anomaly cancellation equations give (see \eqref{sec:gaugeanomaly}),
\begin{align}\label{su2example}
\begin{split}
&n_F =- 8K\cdot E_0 - 2E_0^2 = 16-16g+ 6E_0^2,\\
&n_{\text{adj}} = 1 + {1\over 2} \left(K\cdot  E_0 + E_0^2 \right)=g,
\end{split}
\end{align}
where we have used the adjunction formula $-K\cdot E_0-E_0^2 =2-2g$. 

Finally in Step 3, we have
\begin{align}\label{su2eg2}
{1\over 2} \left(
\sum_{\alpha\in G} |\varphi(\alpha)|^3 -\sum_\mathbf{R} n_\mathbf{R} \sum_{w\in \mathbf{R}} | \varphi(w)|^3\right) = (8-8n_{\text{adj}} - n_F) |\varphi_1|^3,
\end{align}
where we used the $\varphi_i$-basis in \eqref{varphibasis}. The evaluations of the simple coroot $D_1$ on the positive root $\varepsilon_1$ and the highest weight $\sigma_1$ in the fundamental representation are normalized as $D_1(\varepsilon_1)=2$ and $D_1(\sigma_1) = 1$. Plugging \eqref{su2example} into the cubic coefficient in \eqref{su2eg2}, we obtain the triple intersection number of $D_1$,
\begin{align}
D_1^3= -8+8g-6E_0^2,
\end{align}
which agrees with the geometric calculation \eqref{su2triple}.

\section*{Acknowledgements} 
We thank Paolo Aluffi, Clay C\'ordova, Michele Del Zotto, Thomas  Dumitrescu, Babak Haghighat, Ben Heidenreich, Simeon Hellerman, Kentaro Hori, Monica Jinwoo Kang, Guglielmo Lockhart, Tom Rudelius, Cumrun Vafa, Dan Xie, Masahito Yamazaki, Shing-Tung Yau, Xi Yin, and especially to Wati Taylor for many interesting discussions. ME thanks Dartmouth College and The City University of New York for their hospitality.  SHS would like to thank the Kavli Institute for the Physics and Mathematics of the Universe (IPMU) and National Taiwan University for their hospitality. ME is supported in part by the National Science Foundation (NSF) grant DMS-1406925 ``Elliptic Fibrations and String Theory". The work of SHS is supported by a Kao Fellowship at Harvard University.

\appendix

\section{Triple Intersection Numbers in the Resolved I$_N^s$ Model}\label{app:intersection}

In this appendix we present the detailed calculation of the triple intersection numbers in the resolved Weierstrass model of type I$_N^s$. Our method is completely general and can be applied to other Weierstrass models.

 In \S \ref{app:pushforward} we review the definitions of a pushforward and a pullback map \cite{fultonintersection}. We state two theorems that will be important  for computing the triple intersection numbers. In \S \ref{app:su2}, we compute the triple intersection numbers in the simplest nontrivial example, the resolved I$_2^s$ Weierstrass model. 
 In \S\ref{app:sun} we generalize the calculation to one particular resolution $\mathscr{T}$ of the I$_N^s$ model defined in  \S\ref{sec:INmodel}.

\subsection{Pushforwards and Pullbacks}\label{app:pushforward}

Let  $f: M\rightarrow N$ be a proper morphism. Let $[C]$ be a class in the Chow ring of $N$ and $[D]$ be a  class in the Chow ring of $M$. The \textit{pushforward} $f_*$ is defined as
\begin{align}
f_*[D] = \begin{cases}
&0 ~~~~~~~\,~~~~~~~~~~\,~~~\text{if  } \dim \, f(D)\neq \dim \, D,\\
&d \, [f(D)] ~~~~~~~~~~~\text{if  } \dim \, f(D)= \dim \, D,
\end{cases}
\end{align}
where $d$ is the number of points (counting with multiplicities) of the preimage $f^{-1}(y)$ of a generic point $y\in f(D)$.\footnote{To be more precise, $d$ is the degree of the field extension $[R[D]:R[f(D)]]$ where $R[D]$ denotes the field of rational function of the variety $D$.}

We will define the \textit{pullback} $f^*$ for a flat morphism $f$, \textit{i.e.} a morphism whose preimage  is equidimensional,   as
\begin{align}
f^* [C] = [ f^{-1}(C) ].
\end{align}

The intersection product satisfies the \textit{projection formula} (see, for example, \cite{fultonintersection}):
\begin{align}\label{axiom}
f_*( f^* [C] \cdot [D] ) = [C]\cdot f_* [D].
\end{align}
From now on we will drop the bracket for the class $[C]$ when there is no potential confusion.

Two central theorems for the pushforward maps are stated below. The pushforward of the Chern class and  the exceptional divisor is given by the following theorem \cite{aluffi2010chern,fullwood2012stringy}:

\paragraph{Theorem 1}
Let $f:\tilde X\rightarrow X$ be the blowup of a smooth variety $X$ along a smooth complete intersection $V:(F_1=F_2=\cdots=F_k=0)\subset X$, let $E$ be the class of the exceptional divisor in $\tilde X$ and let $U_i$ be the class of $F_i=0$ in $X$. Then
\begin{align}\label{thm3}
c(\tilde X) = { (1+E) (1+f^* U_1 - E) \cdots (1+f^* U_k -E)\over (1+f^* U_1)\cdots (1+f^*U_k)} ~f^* c(X),
\end{align}
and
\begin{align}\label{thm1}
f_\ast (E^n) = \sum_{\alpha=1}^k \left(  \prod_{\beta\neq \alpha} {U_\beta\over U_\beta-U_\alpha} \right)(U_\alpha)^n.
\end{align}
Or equivalently,
\begin{align}
f_*(E-E^2+E^3-\cdots)  =  \prod_{\alpha=1}^k{U_\alpha\over 1+U_\alpha}.
\end{align}

The pushforward via a projection map $\pi$ of a projective bundle is given by (see, for example, \cite{fultonintersection}):
\paragraph{Theorem 2}

Let $V$ be a vector bundle over $B$ and $\pi: \mathbb{P}(V)\rightarrow B$ be its projectivization. Let the class for the canonical line bundle $\mathscr{O}(1)$ over $\mathbb{P}(V)$ be $H$. Then
\begin{align}\label{thm2}
\pi_* \left({1\over 1- H}\right) = {1\over c(V)}.
\end{align}

By applying the above two theorems  as well as the projection formula \eqref{axiom} repeatedly, we can map the classes in the resolved space to  the classes on the base $B$. Hence the final triple intersection numbers are expressed in terms of the intersection data in $B$.

\subsection{The I$_2^s$ Model}\label{app:su2}

We begin with the resolved Weierstrass model obtained in  \cite{Esole:2014bka} with I$_2^s$ singular fiber. Since the singular Weierstrass model $\mathscr{E}_0$ is the zero locus of a section of the bundle $\mathscr{O}(3)\oplus \pi^*\mathscr{L}^6$ over $Y_0$, its class $[\mathscr{E}_0]$ can be expressed as\footnote{Let $C$ be a class in a variety $N$. Whenever there are potential confusions about which variety (or to be precise, which Chow ring) the class $C$ belongs to, we will cap it with the class of the ambient space $[N]$, \textit{i.e.} $C\cap [N]$.
}
\begin{align}
[\mathscr{E}_0] =(3H + 6\pi^* L)\cap [Y_0].
\end{align}
Here $H$ is the class of the canonical line bundle $\mathscr{O}(1)$ over the projective bundle $Y_0=\mathbb{P}(\mathscr{O}_B\oplus \mathscr{L}^2\oplus \mathscr{L}^3)$. 

 The resolve space is obtained by a single blowup $f_1$ with center 
\begin{align}
(x,y,e_0).
\end{align}
The network of blowups in this case is
\begin{align}\label{I2blowup}
\mathscr{E}_0 \xleftarrow{(x,y,e_0|e_1)} \mathscr{T}
\end{align}
where $E_1:~e_1=0$ is the exceptional divisor. Denote the ambient fourfold obtained from blowing up $Y_0$ by $Y_1$. We have the following diagram for the ambient fourfolds $Y_i$ and the base:
\begin{align}
\begin{split}
&Y_0 \xleftarrow{~f_1~} Y_1 \\
&\big\downarrow\,\pi\\
&B
\end{split}
\end{align}
After factoring out the exceptional divisor class $E_1$ from the total transform, the class of resolved space $[\mathscr{T}]$ in  $Y_1$ is \cite{Esole:2014bka}
\begin{align}
[\mathscr{T}] =( 3f_1^\ast H+6f_1^\ast \pi^\ast L -2E_1)\cap [Y_1].
\end{align}

The class for the surface $D_1$  swept out by the node in the ${su}(2)$ Dynkin diagram along the curve $E_0$ is simply the class of the exceptional divisor\footnote{We will ignore   $D_0$  from now on.}
\begin{align}
D_1=E_1\cap [\mathscr{T}].
\end{align}
Note that the class $E_1$ is in the resolved Weierstrass model $\mathscr{T}$ rather than the ambient fourfold $Y_1$ obtained by blowing up $Y_0$. 
To compute the triple intersection number $D_1\cdot D_1\cdot D_1$, we need to pushforward powers of $E_1$ by $f_{1*}$ to $Y_0$ first, and then by $\pi_*$ to the base $B$.

To begin with, we compute the pushforward of $E_1^n$ under the first blowup map $f_1$ using Theorem 1. The classes for the center of the first blowup are
\begin{align}
U_1= H +2\pi^* L,~~U_2=H+3\pi^\ast L,~~U_3= \pi^\ast E_0.
\end{align}
Applying Theorem 1, we then obtain the pushforward classes in $Y_0$. We record $f_{1*}(E_1^n)$ up to $n=4$ below, which are relevant to our calculation of the triple intersection:
\begin{align}\label{pushf1}
\begin{split}
&f_{1*}1=1,~~~~~~~~~f_{1*}(E_1)=0,~~~~~~~~~f_{1\ast} (E_1^2)=0,\\
&f_{1*} (E_1^3)= \pi^\ast E_0 (H+2\pi^\ast L)(H+3\pi^\ast L),\\
&f_{1*} (E_1^4)= \pi^\ast E_0 (H+2\pi^\ast L)(H+3\pi^\ast L)(\pi^\ast E_0+H+5\pi^\ast L).
\end{split}
\end{align}

Next, we apply Theorem 2 to compute the pushforward of the class $H$ to the base $B$:
\begin{align}
{1\over (1+2L)(1+3L)} = \pi_\ast \left({1\over 1-H}\right).
\end{align}
The ambient projective bundle $Y_0$ is a fourfold while the singular Weierstrass model is a hypersurface in $Y_0$ cut out by a section of $\mathscr{O}(3)\oplus\pi^*\mathscr{L}^6$. The relative dimension of the projection map $\pi$ is therefore 2. By matching the dimension, we have
\begin{align}\label{su2pushpi}
\pi_\ast H=0,~~\pi_\ast H^2=1,~~\pi_\ast H^3 =-5L.
\end{align}

Combining \eqref{pushf1} with \eqref{su2pushpi} (as well as the projection formula  \eqref{axiom}), we obtain the pushforward of the class $E_1^4$, $(E_1^3 \cdot f_1^* H)$, and $(E_1^3 \cdot f_1^*\pi^*L)$ to the base $B$,
\begin{align}
\pi_* f_{1*} E_1^4= E_0  (E_0+5L),~~~\pi_* f_{1*} \left(  E_1^3\cdot f_1^*H\right)=0,~~~\pi_* f_{1*}\left(  E_1^3\cdot f_1^*\pi^* L\right)=E_0L.
\end{align}
The triple intersection  $D_1\cdot D_1\cdot D_1$ is then simply a linear combination of the above three quantities, 
\begin{align}
\begin{split}
D_1^3&= \pi_* f_{1*} (D_1^3)=
 \pi_* f_{1*} (E_1^3\cap [\mathscr{E}_1]) \\
 &=\pi_*f_{1*}\Big(  E_1^3\cdot( 3f_1^\ast H+6f_1^\ast \pi^\ast L -2E_1) \cap [Y_1]\Big)\\
 &=-2E_0^2 -4E_0\cdot L.
 \end{split}
 \end{align}

In particular, in the case when the Weierstrass model is Calabi-Yau, we have $ L=-K$ and by the adjunction formula, $-K\cdot E_0 = E_0^2 +2-2g$. It follows that
\begin{align}\label{tripleint2}
D_1^3= -8+8g -6E_0^2.
\end{align}
To match with the 5$d$ $su(2)$ gauge theory prepotential \eqref{E1}, we first note that the number of adjoint hypermultiplets is the genus of the curve $E_0$, 
\begin{align}
n_{\text{adj}}=g.
\end{align}
This can be seen from the 4-supercharge quantum mechanics from a wrapped M2-brane \cite{Witten:1996qb} (see also \cite{Katz:1996ht}). We then have $D_1^3 = 8-8g-n_F$, which determines the number of fundamental hypermultiplet:
\begin{align}
n_F = 16-16g+6E_0^2.
\end{align}
Note that there is no bare Chern-Simons level in the $su(2)$ case because there is no nontrivial third-order Casimir.

\subsection{The I$_N^s$ Model}\label{app:sun}

Given a singular Weierstrass model  with a singular fiber of type I$_N^s$, there are many different small resolutions. The triple intersection numbers are different for each resolution. In this paper we concentrate on one particular resolution $\mathscr{T}$ defined in \S\ref{sec:INmodel}. For $N=2n$, the sequence of blowups for $\mathscr{T}$ is
\begin{align}\label{rI2n}
\begin{split}
\mathscr{T} : ~\mathscr{E}_0 \, \underset{f_1}{\xleftarrow{(x,y,e_0|e_1) } }
\, \mathscr{E}_1\,\underset{f_2}{ \xleftarrow{(y,e_1|e_2) }}
 \,\mathscr{E}_2 \,\underset{f_3}{\xleftarrow{(x,e_2|e_3)} }
  \cdots
 \,\underset{f_{2n-1}}{ \xleftarrow{(x,e_{2n-2}|e_{2n-1})}}\,
  \mathscr{T}
  \end{split}
\end{align}
For $N=2n+1$, the sequence of blowups for $\mathscr{T}$ is
\begin{align}\label{rI2n1}
\begin{split}
\mathscr{T} : ~\mathscr{E}_0 \,\underset{f_1}{\xleftarrow{(x,y,e_0|e_1) }}
\,\mathscr{E}_1\,\underset{f_2}{ \xleftarrow{(y,e_1|e_2) }}
 \,\mathscr{E}_2\, \underset{f_3}{\xleftarrow{(x,e_2|e_3)} }
\,\cdots \, \underset{f_{2n}}{\xleftarrow{(y,e_{2n-1}|e_{2n})}}
 \, \mathscr{T}
  \end{split}
\end{align}
The variables in the parentheses are the center of the blowup. With the exception of the first blowup $f_1$, the center is alternating between $(x,e_i)$ or $(y,e_i)$, where $e_i=0$ being the exceptional divisor.  

The singular Weierstrass model $\mathscr{E}_0$ is defined as a hypersurface in an ambient projective bundle $Y_0=\mathbb{P}(\mathscr{O}_B \oplus \mathscr{L}^2\oplus \mathscr{L}^3)$. We will denote the ambient fourfold after the $i$-th blowup by $Y_i$. We have the following diagram for the ambient fourfolds and the base:
\begin{align}
\begin{split}
&Y_0 \xleftarrow{~f_1~} Y_1 \xleftarrow{~f_2~}\cdots \xleftarrow{f_{N-1} } Y_{N-1}\\
&\big\downarrow\,\pi\\
&B
\end{split}
\end{align}
The resolved Weierstrass model $\mathscr{T}$ is a hypersurface in $Y_{N-1}$ with class\footnote{To simplify the notations, we will not write the pullback map $f_i^*$ or $\pi^*$ explicitly in this subsection.}
\begin{align}
[\mathscr{T}]= (3  H + 6  L -2E_1  - \sum_{i=2}^{N-1} \, E_i  )\cap [Y_{N-1}].
\end{align}
The classes for the surfaces $D_i$  swept out by the $i$-th node can be determined similarly as in the lower rank cases:
\begin{align}\label{rCclass}
D_i = 
\begin{cases}
\,E_{2i-1} -E_{2i},~~\,~~~\,~~~~~~~\text{if}~~i<\left\lceil {N\over 2}\right\rceil ,\\
\, E_{N-1},~~~~~~~~~~~\,~~\,~~~~~~\text{if}~~i=\left\lceil {N\over 2}\right\rceil,\\
\,E_{2N-2i} -E_{2N-2i+1},~\,~~~\text{if}~~i>\left\lceil {N\over 2}\right\rceil.
\end{cases}
\end{align}
understood as classes in the threefold $\mathscr{T}$. The singular fibers for the resolved I$_N^s$ model is shown in Figure \ref{fig:IN}.

To compute the triple intersection number $D_i \cdot D_j \cdot D_k \cap [\mathscr{T}]$, we first determine the pushforward of the exceptional divisor $E_i$ under $f_{i*}$. This is given by \eqref{thm1}
\begin{align}\label{pushf}
f_{i*}(E_i^n) =  \sum_{\alpha=1}^k \,
\left(\prod_{\substack{\beta=1\\\beta\neq \alpha}}^k {U_\beta^{(i)} \over U_\beta^{(i)} -U_\alpha^{(i)} }\right) \, \left( U_\alpha^{(i)} \right)^n .
\end{align}
 with the centers of the blowups given by:\footnote{For $i=2$ and $i=3$ we have  $U_1^{(i=2)} = H-3K-E_1$ and $U_1^{(i=3)} = H-2K-E_1$.}
\begin{align}\label{center}
\{ U_\alpha^{(i)} \}  = 
\begin{cases}
~U_1^{(i)}= H+2L , ~~~~~U_2^{(i)}=H+3L ,\,~~~~~U_3^{(i)}= E_0 ~~~~~~\text{if}~~i=1~~~~~~~~~~~~~~(k=3),\\
~U_1^{(i)}=  H+3L -E_1 -\sum_{j=1}^{{i-2\over2} }E_{2j} ,~~~~~~~~~U_2^{(i)}= E_{i-1}  ~~~~\text{if}~~i:\text{even}~~~~~~~~~~~(k=2),\\
~U_1^{(i)}=  H+2L -E_1 -\sum_{j=1}^{{i-3\over2} }E_{2j+1} ,~\,~~~~~U_2^{(i)}= E_{i-1}  ~~~~\text{if}~~i:\text{odd,}~i\ge3~\,~~~(k=2).
\end{cases}
\end{align}
Here $k$ is the number of generators in the center of the blowup.

 Next, we compute the pushforward of $H^n$ under the projection map $\pi$. This is given by Theorem 2 with the choice $c(V) = (1+2L)(1+3L)$,
\begin{align}\label{pushpi}
\begin{split}
&\pi_*1=0,~~~~~~~\pi_* H =0,\\
&\pi_* (H^{n+2})  = \left[\, {1\over n!} {d^n\over dL^n} {1\over (1+2L)(1+3L)} \,\right]_{L=0}L^{n}.
\end{split}
\end{align}

With the above preparation, we can express the triple intersection number $D_i\cdot D_j \cdot D_k \cap [\mathscr{T}]$ as
\begin{align}\label{finalint}
D_i\cdot D_j \cdot D_k = \pi_* \circ f_{1*} \circ \cdots \circ f_{N-1*} \left[ \,
D_i\,  D_j\, D_k \,(3  H + 6  L -2E_1  - \sum_{i=2}^{N-1} \, E_i  )\, \right]
\end{align}
with $D_i$ given by \eqref{rCclass} and $f_{i*}$ and $\pi_*$ defined in \eqref{pushf} and \eqref{pushpi}. In the case when the Weierstrass model is Calabi-Yau, we further choose $L=-K$. We list the triple interaction forms for small $N$ in Table \ref{table:intersection}.

\section{The Euler Characteristic of  the Resolved I$_N^s$ Model}\label{app:Euler}

In this appendix we compute the Euler characteristic of the resolved Weierstrass model with a singular fiber of type I$_N^s$ over an arbitrary algebraic surface $B$.  Since all small resolutions of a singular Weierstrass model have the same Hodge numbers \cite{batyrev}, we can choose one particular small resolution and compute its Euler characteristic. For concreteness, we focus on the resolution $\mathscr{T}$ for the I$_N^s$ model defined in \S\ref{sec:INmodel}.

To compute the Euler characteristic, we need to know the Chern classes of the resolved Weierstrass model. These can be systematically determined by Theorem 1. By applying \eqref{thm3} repeatedly, we  can express the Euler characteristic $\chi = \int_{ \mathscr{T}} c(\mathscr{T})$ in terms of the intersection data in the base $B$.

Explicitly, we can express the Chern class of the ambient fourfold $Y_{N-1}$ as\footnote{To simplify the notations, we will not write the pullback map $f_i^*$ or $\pi^*$ explicitly from now on in this subsection.}
\begin{align}
c(Y_{N-1})  = \left(\prod_{i=1}^{N-1}  {(1+E_i) (1+U_1^{(i)} - E_i ) \cdots (1+ U_k^{(i)}-E_i)\over (1+U_1^{(i)}  ) \cdots (1+ U_k^{(i)})} \right) c(Y_0),
\end{align}
with the center of the blowup $U_\alpha^{(i)}$ given in \eqref{center}. The Chern class for the projective bundle $Y_0 = \mathbb{P}( \mathscr{O}_B\oplus \mathscr{L}^2 \oplus \mathscr{L}^3)$ is
\begin{align}
c(Y_0 ) =(1+H) (1+ H+2L )(1+H+3L)c(B).
\end{align}
Using the adjunction formula, the Chern class of the resolved  I$_N^s$ model is 
\begin{align}
c(\mathscr{T}) = { c(Y_{N-1}) \over 1+ 3H+6L-2E_1 -\sum_{i=2}^{N-1}E_i}.
\end{align}
The Euler characteristic is the integral of $c(\mathscr{T})$ on $\mathscr{T}$, which is
\begin{align}
\chi = \int_{\mathscr{T}}c(\mathscr{T}) =c_3(\mathscr{T}) \cdot (3H+6L-2E_1 -\sum_{i=2}^{N-1}E_i) \cap [Y_{N-1}].
\end{align}
Finally, we pushforward the calculation down to the base by applying \eqref{pushf} and \eqref{pushpi} repeatedly,
\begin{align}\label{preeuler}
\chi= \pi_* \circ f_{1*} \circ \cdots \circ f_{N-1*} \,\left[ \,
c_3(\mathscr{T}) \cdot (3H+6L-2E_1 -\sum_{i=2}^{N-1}E_i)\,\right],
\end{align}
with $f_{i*}$ and $\pi_*$ defined in \eqref{pushf} and \eqref{pushpi}. 
We list the Euler characteristics for the resolved I$_N^s$ for small $N$ in Table \ref{table:euler} in the Calabi-Yau case $L=-K$. From direct inspections on the answers for various values of $N$, we obtain a closed form formula in \eqref{euler}.

\section{Notations}

As the calculations of the triple intersection number and the Euler characteristic are quite notationally intense, we summarized the definitions of various symbols here.

\begin{itemize}
\item $B$: ~the base complex surface.
\item $L =c_1(\mathscr{L})$:~the class for the line bundle $\mathscr{L}$ over the base $B$. In the Calabi-Yau case, $L$ is the anticanonical class of the base, $L=-K$.
\item $E_0:$~the class for the divisor $e_0=0$ in the base $B$ that supports singular fibers.
\item $V=\mathscr{O}_B\oplus \mathscr{L}^2\oplus \mathscr{L}^3$.
\item $\pi: Y_0=\mathbb{P}(V)\rightarrow B$:~ the ambient projective bundle. The singular Weierstrass model is a hypersurface in $Y_0$.
\item $\mathscr{O}(1)$:~the canonical line bundle over $Y_0$. $H=c_1(\mathscr{O}(1))$.
\item $[\mathscr{E}_0]= 3H+ 6\pi^\ast L$:~\text{the class of the singular Weierstrass model in $Y_0$.}
\item $Y_i$:~\text{the ambient space for the Weierstrass model obtained after the $i$-th blowup}. 
\item $f_i:Y_i\xrightarrow{~} Y_{i-1}$:~ \text{the $i$-th blowup map}.
\item $E_i$:~\text{the exceptional divisor for the $i$-th blowup in $Y_i$}.
\end{itemize}

\bibliography{tripleint}
\bibliographystyle{utphys}
\end{document}